\documentclass[11pt,a4paper]{article}
\usepackage[left=2cm ,top=25mm,bottom=25mm,right=2cm]{geometry}
\usepackage{tikz}
\usepackage[utf8]{inputenc}
\usepackage{amsmath,amssymb}
\usepackage{graphicx}
\usepackage{float}
\usepackage{graphicx} 
 \usepackage{multirow} 
\usepackage{hyperref}

\usepackage{subfigure}
\usepackage{mathtools}
\usepackage{pdfpages}
\numberwithin{equation}{section}

\usepackage{orcidlink}
\usepackage{cite}

\begin{document}

\thispagestyle{empty}
\begin{titlepage}
	\vspace*{0.7cm}
	\begin{center}
		{\Large {\bf String loop origin for dark radiation and superheavy dark matter in type IIB compactifications}}
		\\[12mm]
		Vasileios Basiouris~$^{\orcidlink{0000-0002-9266-7851}}$~\footnote{E-mail: \texttt{v.basiouris@gmail.com}}
	\end{center}
	\vspace*{0.50cm}
	\centerline{\it
		Physics Department, University of Ioannina}
	\centerline{\it 45110, Ioannina, 	Greece}
	\vspace*{1.20cm}
	
\begin{abstract}
In this article we study the significance of string loop corrections, in a perturbative moduli stabilization scenario, focusing on their role in unraveling the origin of dark radiation in the late cosmological epoch and their correlation to dark matter. More specifically, a detailed analysis is provided in which the mass hierarchy of the normalized fields in the K{\"a}hler moduli sector is determined by the scales of the integer fluxes and the quantum corrections. Furthermore, we compute the previously underestimated contributions to the decay rates of moduli to axions, which behave as dark radiation, highlighting their connection to the aforementioned higher-order corrections. Two contrasting reheating scenarios (low-scale and high-scale) are provided, depending on the decay rate of the longest-lived particle into the Standard Model degrees of freedom through a Giudice-Masiero mechanism, while the effective number of neutrino species $\Delta N_{eff}$ remains below the respective bounds. Finally, a non-thermal dark matter scenario is proposed based on the decays of heavy scalar fields, where the main production mechanisms are investigated, leading to a dark matter candidate mass ranging from a few $GeV$ up to $10^{12}\; GeV$.\vspace{2cm}

\end{abstract}
\end{titlepage}

\section{Introduction}

Developments on the attainability of de-Sitter (dS) vacuum in type IIB string compactifications have been widely discussed in the literature recently. Despite the proposed swampland conjectures \cite{hep-th/0301240, hep-th/0509212, hep-th/0605264, 1610.01533, 1806.08362,  1810.05506, 1806.09718, 1807.06581, 2008.13251}, several works point toward a possibility of a dS vacuum either by incorporating non-perturbative corrections in the superpotential \cite{hep-th/0409215,hep-th/0502058,hep-th/0505076,0708.1873, 0802.1557, hep-th/0501032,2007.04327} or by including perturbative quantum dynamics in the K{\"a}hler potential \cite{1803.08941,1909.10525,2203.03362,2007.15423,2109.08421}. Among the most plausible explanations for the aforementioned question are focusing on the study of the effective theory, where an Anti de-Sitter (AdS) vacuum is evident and various uplifting ingredients are included ($\bar{D}_3$ branes and D-terms) in order to achieve a dS vacuum. The central role in these setups is played by the moduli fields of the theory, which modify not only the relevant scales for the correct embedding of inflation but also the late-time cosmological dynamics of the universe such as the reheating temperature and the effective neutrino species number. A potential connection between dark matter and dark radiation can also be implemented in these frameworks due to the decays of moduli to particles of the dark sector. Some recent references regarding the open problem of moduli stabilization, dark radiation and their correlation to dark matter can be found in the following works \cite{2310.20559,2303.04819, Chauhan:2025rdj, Ebelt:2023clh, 2008.10625,1912.10047,hep-th/0502058,hep-th/0505076, 1304.0792, Ling:2025nlw, Manno:2025dhw, Sheridan:2024vtt}.\par 
In the present paper, we focus on the importance of quantum string corrections to the K{\"a}hler potential, whose origin can be traced back to the higher derivative terms of the effective string action and the existence of localized Einstein-Hilbert terms \cite{hep-th/9812093,hep-th/0209030,hep-th/9707013,hep-th/9608012} in a geometric setup of three intersecting $D_7$ branes. Their inclusion in the theory provides a novel way to stabilize the K{\"a}hler moduli fields of the theory without considering non-perturbative corrections, whose dynamics could be problematic regarding the value of the string coupling in different parametric regions of the theory. Moreover, the dS vacuum is achieved by considering the presence of magnetic fluxes along the cycles of the $D_7$ branes, which induce anomalous $U(1)$ symmetries, thereby charging the K{\"a}hler moduli fields. Their result is an induced D-term at the level of the effective potential \cite{hep-th/0309187,hep-th/0602253}, which despite being moduli dependent, could in principle suffice to uplift the Anti-de Sitter vacuum.

In a second stage, this work focuses on a detailed calculation of the moduli mass eigenstates and eigenvalues, as well as their correlation with the choice of fluxes $\mathcal{W}_0$ (either exponentially suppressed or of order one), which characterize the mass hierarchy and the potential determination of the longest-lived particle field. Based on this, the couplings of the normalized fields to the axions, which are embedded in the dark sector of the theory, are computed unraveling that not only the diagonal decays of the moduli to axions are important, but also that off-diagonal contributions in the K{\"a}hler metric contribute substantially to the dark radiation abundance. On the other hand, the dominant decay in the visible sector is described by the Giudice-Masiero mechanism \cite{Giudice:1988yz}. It is calculated for the two limiting cases with respect to the scale of the fluxes $\mathcal{W}_0$, leading to a theoretical estimation of the effective number of neutrino species $\Delta N_{eff}$. Additionally, a distinction is provided regarding the relevant scale of the reheating temperature ($\mathcal{O}(\text{MeV})\leq T_{rh} \leq \mathcal{O}(\text{GeV})$ scale for exponentially suppressed fluxes and $T_{rh}\gg \mathcal{O}(\text{Tev})$ for order one fluxes), where this explanation can be correlated with an early matter-dominated phase as noted in \cite{2203.08833}.

Finally, we study the production of non-thermal dark matter after the reheating process. The primary candidates in large volume limit scenarios are the weakly interacting massive particles (WIMPs) and the thermally underproduced (Higgsino-like or Bino-like) particles, which in many previous studies tend to be overproduced \cite{2208.01017,WileyDeal:2025wgh,1208.3562}. An alternative scenario proposes the fuzzy dark matter case, where axions play a central role in this approach \cite{2110.02964,2312.13431}. In view of the new quantum corrections in the present setup, we study the most common mechanisms of dark matter production - the annihilation scenario and the branching scenario- without the issue of overproduction. We provide a scenario where a possible superheavy dark matter could emerge from the annihilation scenario, with its mass lying at $10^{11} \; \text{GeV}$. In contrast, the branching scenario could also yield the correct dark matter abundance for dark matter particles even below the $\text{TeV}$ scale. Additionally, a low-scale baryogenesis mechanism in \cite{1011.1286,1005.2804,Mohapatra:2025bdl} is employed, making the current approach ideal to furnish an explanation for the dark matter-baryon coincidence. Since the modulus could decay to species with B- and CP- violating couplings with the Standard Model particles and has the correct scaling for the dilution factor of entropy $Y_{\phi}$, it could naturally accommodate this coincidence.\par

Summarizing all the above, in section 2., we provide the preliminary details on the geometric background of the proposed model. We sketch the procedure of moduli stabilization and characterize the properties of the dS vacuum along with the corresponding minima. In section 3., a discussion is presented with respect to the mixing of normalized fields, where the eigenvalues are correlated to the quantum effects and the mass hierarchy is associated to the scale of integer fluxes. In addition, in section 4. we proceed to the main consequences of our hypothesis, computing the reheating temperature and the effective number of neutrino species $\Delta N_{eff}$. These quantities depend on the decay rates of the longest-lived particle to axions and Higgses, resulting in two different scales of reheating temperature. Finally in section 5., we study a dark matter scenario where the two production mechanisms are discussed resulting in two contrasting possibilities for the dark matter mass in accordance with the relevant scale of reheating.

\section{Structure of the Potential}
To set the stage, we begin with the $\mathcal{N}=1$ supergravity K{\"a}hler potential $\mathcal{K}$, where the geometric configuration consists of of three intersecting $D_7$ branes. The theory contains various scalar fields, but we focus our attention on the complex structure moduli ($z_a, \;a=1,2,3$), the axio-dilaton ($S$) and the K{\"a}hler moduli $T_i,\;i=1,2,3$. Supersymmetric conditions $\mathcal{D}_{z_a}\mathcal{W}=\mathcal{D}_{S}\mathcal{W}=0$ fix both the complex structure moduli and the axio-dilaton, leaving effectively the K{\"a}hler sector completely undetermined. The internal volume of the six dimensional space in the context of IIB string theory is denoted by $\mathcal{V}$, and is expressed in terms of the two-cycles of the theory as:

\begin{equation}
    \mathcal{V}=\dfrac{1}{6}k_{ijk}v^i v^j v^k, \; v^i=-\text{Im}(T^i),
\end{equation}

where the tensor $k_{ijk}$ characterizes the intersection number. A more useful formula for the compactified volume can be expressed in terms of the four-cycles $\tau_i$ of the theory, where they are related to the two-cycles as:

\begin{equation}
    \tau_i=v^j v^k \rightarrow \mathcal{V}=\sqrt{\tau_1\tau_2\tau_3}~.
\end{equation}

We assume that the nonzero classical triple intersection number is $k_{123}=1$. Apart from the compactified volume, the K{\"a}hler potential contains the $(\alpha'^3)$ correction $\xi$, where this correction corresponds to a constant shift of the volume \cite{Becker:2002nn}. In addition, we include the effects of quantum string loop corrections along each world-volume direction of the internal space, incorporated in a perturbative form as $\eta \log(\tau_i)$ \cite{1909.10525}. Their origin can be traced back to the higher-derivative terms of the 10-dimensional supergravity theory, where the leading effects appear as a $\mathcal{R}^4$ term, with $\mathcal{R}$ being the Riemann curvature. After dimensional reduction to four dimensions, these effects induce a localized Einstein-Hilbert term, where the computation of the scattering amplitude between these localized graviton vertices and $D_7$ branes (in the form of closed string modes) leads to a perturbative form of the correction $\eta \log(\tau_i)$ at the K{\"a}hler potential level \cite{1909.10525}.

\begin{equation}
\mathcal{K}=-2\log(\sqrt{\tau_1\tau_2\tau_3}+\xi+\eta_i\log(\tau_i)),\quad  \eta =-\frac{1}{2}g_s T_i \xi~.\label{Kahler}
\end{equation}

For simplicity, we assume the perturbative parameter $\eta$ to be identical along each direction ($\eta_1\cong \eta_2 \cong \eta_3$), i.e. the string tension $T_i$ of the corresponding branes is tuned to be the same. Regarding the superpotential of the theory, we assume the existence of background fluxes $\mathcal{W}_0$ \cite{Gukov:1999ya} and the non-perturbative effects are turned off.
\begin{equation}
\mathcal{W}=\mathcal{W}_0~.\label{fluxes}
\end{equation}

The computation of the F-term potential is completely straightforward, taking into account equations \eqref{Kahler}, \eqref{fluxes} and trading one modulus, e.g. $\tau_3=\mathcal{V}/(\tau_1\tau_2)$, the whole effective potential is expressed in terms of the volume:
\begin{equation}
V_F=\frac{3 \mathcal{W}_0^2 (-8 \eta +\xi +2 \eta  \log (\mathcal{V}))}{2\mathcal{V}^3}-\frac{9 \;\eta  \;\xi  \mathcal{W}_0^2 \log (\mathcal{V})}{\mathcal{V}^4}+\mathcal{O}(\frac{1}{\mathcal{V}^n})~.\label{Fpotential}
\end{equation}

It is important to highlight the fact that this very compact and illustrative formula has been obtained, considering that we would like to study the large volume limit where quantum corrections are subleading. This fact allows us to perform an expansion in terms of $\eta$ and $\frac{\xi}{\mathcal{V}^n}$, while the terms proportional to the power of the expansion variables are neglected. In addition, since the leading order terms are of order $\sim \mathcal{O}(\dfrac{1}{\mathcal{V}^3})$, we do not consider terms of order bigger than $\sim \mathcal{O}(\dfrac{1}{\mathcal{V}^4})$ in the large volume regime, bearing in mind that these additional terms are proportional to powers of $\eta$ making them less important. In order to obtain an AdS vacuum, we have to compute the minimum along the volume $\mathcal{V}$ direction, which is given by:

\begin{equation}
\mathcal{V}_{min}=e^{\frac{13}{3}-\frac{\xi }{2 \eta }}~.
\end{equation}

The uplift mechanism for realizing a dS minimum is accomplished by adding the D-terms, related to the three intersecting D7-branes of the geometric configuration. Flux generated D-terms \cite{hep-th/0309187, 2203.03362, Haack:2006cy, Cicoli:2011yh} have the following form:

\begin{equation}
    V_D=  \sum_{i=1}^3 \dfrac{g^2_{D_{7_i}}}{2} (\sum_{i\neq j} Q_{ij} \partial_{T_j} \mathcal{K} + \sum_{j \neq i}q_i^j |\Phi_i^j|^2)^2, \label{GenD}
\end{equation}

where $g_{D_{7_i}}$ stands for the gauge coupling of the $D_7$ brane, $Q_{ij}$ represents the charges of the K{\"a}hler moduli, while $q_i^j, \Phi_i^j$ are the charges and the scalar components of the superfields, respectively. Considering that the vevs of the matter fields are $\langle \Phi_i^j \rangle = 0$, the formula is significantly simplified to:

\begin{equation}
    V_D \cong \sum_{i=3}^3 
    \big[\dfrac{1}{\tau_i}\big(\sum_{i\neq j}Q_{ij} \partial_{T_j} \mathcal{K}\big)^2]\cong \sum_{i=1}^3 \dfrac{d_i}{f_a^3},\;\; d_i\cong Q^2_{ij}>0,\label{gd}
\end{equation}

where $f_a^3$ is a cubic polynomial parametrized by a generic four-cycle modulus $\tau_j$. Now, the above formula can be further approximated, as noted in  \cite{2203.03362,Bera:2024zsk, Ahmed:2022dhc}, by considering the toroidal-like symmetry of the underlying geometry. Moreover, the three intersecting stacks of $D_7$ branes (which is the geometric setup of this model) are associated with gauge groups, where in principle the magnetic fluxes can induce some anomalous $U(1)$ symmetries. As rigorously studied in \cite{hep-th/0309187, hep-th/0601190, hep-th/0609211} a suggestion made by Burgess et al. and Achucarro et al.\footnote{A criticism of this approach can be found in \cite{arXiv:hep-th/0503216, arXiv:hep-th/0506266}.}, D-terms of the above form could also be derived from a different origin. From a 4D point of view, this type of D-terms was identified as a Fayet-Iliopoulos term depending on the K{\"a}hler moduli in the $\mathcal{N}=1$ supersymmetric effective action \cite{Dine:1987xk}. Taking into account the anomalous $U(1)$, the four-cycle moduli, parametrizing the transverse volume of the magnetic $D_7$ brane, obtain a charge $Q$ under the $U(1)$. In addition, there are charges $q_i^j$ carried by the scalar fields $\Phi_i^j$, which fields can be minimized to zero. As a consequence of the above discussion, we could write down the D-terms of this intersecting $D_7$ brane model, following the work of \cite{hep-th/0309187, hep-th/0601190, hep-th/0609211, 1803.08941}, as:

\begin{align}
    V_D \cong \sum_{i=1}^3 \dfrac{d_i}{\tau_i^3}~.\label{Dterm}
\end{align}

To support our approach, we provide in Appendix A.3 a detailed proof that the above formula gives an equivalent dS vacuum to the vacuum that can be derived from the generic formula \eqref{gd} up to a rescaling of the uplifting parameters $d_i$. As a consequence, we can argue that our approximation spoils neither the existence of a dS vacuum nor the subsequent analysis. Appending the D-term effects on the F-term potential, the complete effective potential is summarized below: 

\begin{equation}
V_{eff}=\frac{3 \mathcal{W}_0^2 (-8 \eta +\xi +2 \eta  \log (\mathcal{V}))}{2 \mathcal{V}^3}+\frac{d_1}{\tau_1^3}+\frac{d_2}{\tau_2^3}+\frac{d_3}{\tau_3^3}\label{Vfull}~.
\end{equation}

The $\tau_3$ modulus could be traded for the internal volume modulus $\mathcal{V}$, i.e., $\tau_3$ =$\mathcal{V}^2/(\tau_1\tau_2)$. Thus, the effective potential is the sum $V_{eff} = V_F + V_D$, while it can be expressed as a function of $\tau_1,\tau_2$ and $\mathcal{V}$:
\begin{equation}
V_{eff}=\frac{3 \mathcal{W}_0^2 (-8 \eta +\xi +2 \eta  \log (\mathcal{V}))}{2 \mathcal{V}^3}+\frac{d_1}{\tau_1^3}+\frac{d_2}{\tau_2^3}+\frac{d_3 \tau_1^3 \tau_2^3}{\mathcal{V}^6}\label{Veff}~.
\end{equation}

For completeness, we are going to describe the minima along the three transverse directions ($\mathcal{V},\tau_1,\tau_2$). Minimizing along each direction, we get the following minima and some useful relations constraining the free parameters of the theory. Moreover, the potential along the volume direction is displayed below, where the minimal values for the $\tau_1,\tau_2$ moduli have been applied.

\begin{align}
\tau_1=(\dfrac{d_1^2}{d_2d_3})^{1/9}\mathcal{V}^{2/3},\;\;\tau_2=(&\dfrac{d_2^2}{d_1d_3})^{1/9}\mathcal{V}^{2/3},\\
\mathcal{V}_{min}=\frac{3 \eta\;  \mathcal{W}_0^2\; W_{0/-1}\left(\frac{2 d e^{\frac{13}{3}-\frac{\xi }{2 \eta }}}{3 \eta\;  \mathcal{W}_0^2}\right)}{2 d}&,\;\;d=(d_1d_2d_3)^{1/3}\label{mins}~,
\end{align}

\begin{equation}
    V_{eff}(\mathcal{V})=\frac{6 d \mathcal{V}+3 \mathcal{W}_0^2 (\xi -8 \eta )+6 \eta  \mathcal{W}_0^2 \log (\mathcal{V})}{2 \mathcal{V}^3} \label{Potential}~.
\end{equation}

where the $W$-function denotes the Lambert function. The minimum and the maximum along the volume direction are characterized by the upper $W_0$ and the lower branch $W_{-1}$, respectively. Now, a dS minimum puts a stringent bound on the parameter $\rho=\frac{d}{(\mathcal{W}_0)^2},\;d=(d_1d_2d_3)^{1/3}$, where these bounds are obtained from the Lambert's function definition ($W(x),\;x\geq  \frac{1}{e}$) and the positivity of the potential at the minimum.

\begin{align}
-\frac{\eta }{\mathcal{V}}<\rho<-\frac{3}{2} \eta \; e^{\frac{\xi }{2 \eta }-\frac{16}{3}},\;\;\frac{1}{3} \left(26-6 \log \left(-\frac{3 \eta }{2 e \rho}\right)\right)<\frac{\xi }{\eta }<0\label{bound}~.
\end{align}

A different parametrization for the above coefficients is given below, which would be more useful in the following sections:

\begin{align}
    \dfrac{\partial^2 V_{eff}}{(\partial \mathcal{V})^2}=-\dfrac{6d \mathcal{V}_{min}+9\eta \mathcal{W}_0^2}{\mathcal{V}_{min}^5}>0\Rightarrow  \dfrac{\partial^2 V_{eff}}{(\partial \mathcal{V}_{min})^2}=-\dfrac{-9\eta \mathcal{W}_0^2(\frac{2}{3}q+1)}{\mathcal{V}_{min}^5}>0 \Rightarrow q= \dfrac{d \mathcal{V}_{min}}{\eta \mathcal{W}_0^2}>-\frac{3}{2},
\end{align}

\begin{align}
    -\frac{\eta }{\mathcal{V}}<\dfrac{d}{\mathcal{W}_0^2} \Rightarrow q= \dfrac{d \mathcal{V}_{min}}{\eta \mathcal{W}_0^2}<-1,
\end{align}

where by combining the above bounds, the $q$ parameter is strictly bounded between:

\begin{equation}
    -\frac{3}{2}<q<-1~.\label{qpar}
\end{equation}

Clearly, our effective potential could admit a dS vacuum (as depicted in Figure 1.) for various combinations of the parameters either in the exponentially suppressed flux limit or for the order one flux case. 

\begin{figure}[H]
    \centering
    \includegraphics[scale=0.8]{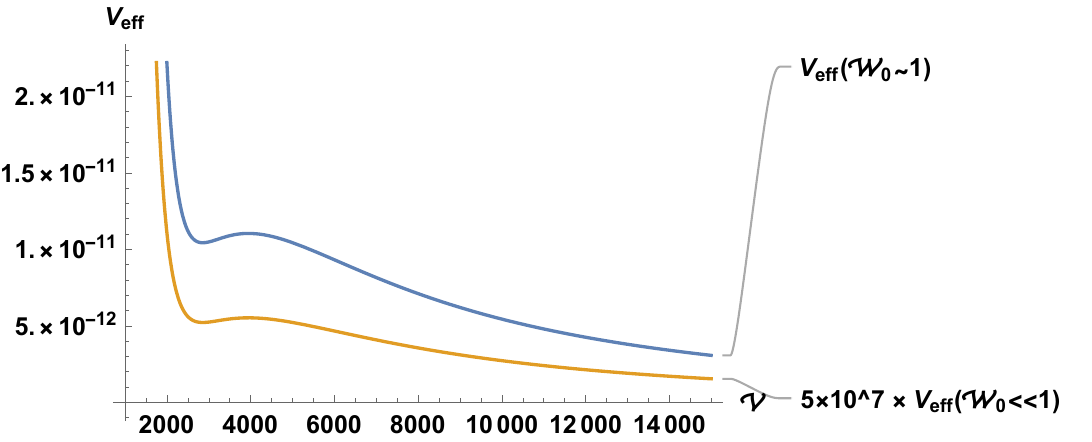}
    \caption{Plot of potential \eqref{Potential} $\xi=5$, $|\mathcal{W}_0|=(1,10^{-4})$, $d=(4\times 10^{-4},4 \times 10^{-12})$ and $\eta=-0.9$.}
\end{figure}

\section{Mixing in the kinetic terms and canonically normalized fields}

Consistently embedding string inflation within type IIB compactifications is one of the challenging problems in studying early universe cosmology. Various works have attempted to study the inflationary evolution of the universe, where different mechanisms are employed \cite{hep-th/0307043, 2208.06606,2305.05703,Ahmed:2022dhc,2009.00653,1810.05060}. The implications of the inflationary scenario in low-energy implications can be viewed indirectly by the correct prediction of the cosmological observables, such as the Big Bang Nucleosynthesis \cite{astro-ph/0601514,astro-ph/0401091}, which is correlated with the scale of inflation and the reheating process. In this section, we are going to present a detailed analysis on the canonical normalization of the moduli fields and highlight the importance of the logarithmic corrections in the off-diagonal entries of the K{\"a}hler metric. In this study, the quantum corrections are of crucial importance, since the eigenvalue of the $\tau_3$ direction is highly influenced by the new effects (parametrized by the quantum parameter $q$), which fact modifies its scaling and mixing with the other sectors. The study of the reheating process and the energy flow to the dark sector is characterized by the decay channels of the longest-lived field, where this will be determined by the diagonalization procedure. The longest-lived particle will dominate the energy density in the late cosmological times and clarify the correlation between different important energy scales in the universe expansion. Following the discussion above, we need to change the basis from the K{\"a}hler moduli $\tau_i$ to the canonically normalized fields $\phi_i$. First of all, following \cite{Cicoli:2010ha}, the mixing of the fields will be characterized by the diagonalization of the mass matrix and the transformation of the basis is driven by the canonically corrected kinetic terms. We start by writing down the definition of the Lagrangian in terms of the moduli fields:

\begin{align}
\mathcal{L}=\mathcal{K}_{ij} \partial_{\mu} \tau_i \partial^{\mu} \tau_j- V-\frac{1}{2} V \tau_i \tau_j +\mathcal{O}(\tau^3),
\end{align}

where $V$ is the scalar potential of the moduli fields, while $K_{ij}$ denotes the K{\"a}hler metric:

\begin{align}
\mathcal{K}_{ij}=\left(
\begin{array}{ccc}
 \frac{1}{4 \tau _1^2} & -\frac{\tau _3 \left(\eta  \log \left(\tau _1 \tau _2 \tau _3\right)+\xi \right)}{8
   \left(\tau _1 \tau _2 \tau _3\right){}^{3/2}} & -\frac{\tau _2 \left(\eta  \log \left(\tau _1 \tau _2
   \tau _3\right)+\xi \right)}{8 \left(\tau _1 \tau _2 \tau _3\right){}^{3/2}} \\
 -\frac{\tau _3 \left(\eta  \log \left(\tau _1 \tau _2 \tau _3\right)+\xi \right)}{8 \left(\tau _1 \tau _2
   \tau _3\right){}^{3/2}} & \frac{1}{4 \tau _2^2} & -\frac{\tau _1 \left(\eta  \log \left(\tau _1 \tau _2
   \tau _3\right)+\xi \right)}{8 \left(\tau _1 \tau _2 \tau _3\right){}^{3/2}} \\
 -\frac{\tau _2 \left(\eta  \log \left(\tau _1 \tau _2 \tau _3\right)+\xi \right)}{8 \left(\tau _1 \tau _2
   \tau _3\right){}^{3/2}} & -\frac{\tau _1 \left(\eta  \log \left(\tau _1 \tau _2 \tau _3\right)+\xi
   \right)}{8 \left(\tau _1 \tau _2 \tau _3\right){}^{3/2}} & \frac{1}{4 \tau _3^2} \\
\end{array}
\right)+\mathcal{O}(\frac{\xi}{\mathcal{V}^n},\eta^n)~.
\end{align}

In the above computation, we have kept the leading order terms ($\sim \mathcal{O}(\eta, \xi/\mathcal{V}^n)$) in the off-diagonal entries, which parametrize the quantum corrections to the kinetic terms. Our main purpose is to observe to what extent these corrections could modify the mixing in the parameter space and their subsequent cosmological implications. The next step would be to compute the mass matrix and the corresponding eigenvalues and eigenvectors. The definition of the mass matrix is given by the following matrix:

\begin{align}
M^2_{ij}=\dfrac{1}{2}(\mathcal{K})^{-1}_{ik}V_{kj},\label{Mass}
\end{align}

where $V_{kj}, i,j=(\tau_1,\tau_2,\tau_3)$ are the second derivatives of the effective potential with respect to the moduli, while the inverse K{\"a}hler metric $\mathcal{K}^{-1}_{ij}$ is given by:

\begin{align}
\mathcal{K}^{-1}_{ij}=\left(
\begin{array}{ccc}
  4 \tau _1^2 & \frac{2 \tau _1 \tau _2 \left(\eta  \log \left(\tau _1 \tau _2 \tau _3\right)+\xi
   \right)}{\sqrt{\tau _1 \tau _2 \tau _3}} & \frac{2 \tau _1 \tau _3 \left(\eta  \log \left(\tau _1 \tau _2
   \tau _3\right)+\xi \right)}{\sqrt{\tau _1 \tau _2 \tau _3}} \\
 \frac{2 \tau _1 \tau _2 \left(\eta  \log \left(\tau _1 \tau _2 \tau _3\right)+\xi \right)}{\sqrt{\tau _1
   \tau _2 \tau _3}} & 4 \tau _2^2 & \frac{2 \tau _2 \tau _3 \left(\eta  \log \left(\tau _1 \tau _2 \tau
   _3\right)+\xi \right)}{\sqrt{\tau _1 \tau _2 \tau _3}} \\
 \frac{2 \tau _1 \tau _3 \left(\eta  \log \left(\tau _1 \tau _2 \tau _3\right)+\xi \right)}{\sqrt{\tau _1
   \tau _2 \tau _3}} & \frac{2 \tau _2 \tau _3 \left(\eta  \log \left(\tau _1 \tau _2 \tau _3\right)+\xi
   \right)}{\sqrt{\tau _1 \tau _2 \tau _3}} & 4 \tau _3^2 \\
\end{array}
\right)+\mathcal{O}(\frac{\xi}{\mathcal{V}^n},\eta^n)~,
\end{align}

\begin{small}
\begin{align}
V_{ij}=\left(
\begin{array}{ccc}
 \frac{12 d_1}{\tau _1^5}+\mu_1 & \tau_3 \lambda & \tau_2 \lambda \\
 \tau_3 \lambda  & \frac{12 d_2}{\tau _2^5}+\mu_2  & \tau_1 \lambda  \\
 \tau_2 \lambda  & \tau_1 \lambda  & \frac{12 d_3}{\tau _3^5}+\mu_3 
\end{array}
\right)~.
\end{align}
\end{small}

\begin{align}
    &\lambda=\frac{9 \mathcal{W}_0^2 \left(3 \eta  \log \left(\tau _1 \tau _2 \tau _3\right)-28 \eta +3 \xi
   \right)}{8 \left(\tau _1 \tau _2 \tau _3\right){}^{5/2}}\\
   &\mu_i=\frac{3 \mathcal{W}_0^2 \left(15 \eta  \log \left(\tau _1
   \tau _2 \tau _3\right)-136 \eta +15 \xi \right)}{8 \tau _i^2 \left(\tau _1 \tau _2 \tau
   _3\right){}^{3/2}}, \quad i=1,2,3~.
\end{align}

Combining all the above ingredients, we can compute the mass matrix of the model as follows:

\begin{small}
\begin{align}
M^2_{ij}\cong\left(
\begin{array}{ccc}
  \frac{24 d_1}{\tau _1^3}+\mu_1  & \tau_3 \tau_1^2 \lambda & \tau_2 \tau_1^2 \lambda \\
 \tau_3 \tau_2^2 \lambda & \frac{24 d_2}{\tau _2^3}+\mu_2  & \tau_1 \tau_2^2 \lambda \\
 \tau_2 \tau_3^2 \lambda & \tau_1 \tau_3^2 \lambda & \frac{24 d_3}{\tau _3^3}+\mu_3 \\
\end{array}
\right)+\; \mathcal{O}(\eta^n,\frac{\xi}{\mathcal{V}^n})~. \label{MassMatrix}
\end{align}
\end{small}

In the above matrix, we have kept the leading order terms taking into account the results of section 2. for the dS vacua, which require $\mathcal{V}\gg 1$. Now, in order to observe the scaling of the moduli at the vacuum, we compute the mass eigenvalues by applying the definitions of equations \eqref{mins} in the trace and the determinant of the mass matrix $M^2$. We expect a similar behavior for the first two moduli $\tau_1, \tau_2$, which makes them heavier than the third modulus. Therefore, we could work the moduli mass spectrum at leading order of volume scaling, noticing that:

\begin{equation}
Tr[M^2]=m_1^2+m_2^2+m_3^2\cong m_1^2 \cong m_2^2,\;\; \dfrac{Det[M^2]}{Tr[M^2]^2}=\dfrac{m_1^2m_2^2m_3^2}{(m_1^2+m_2^2+m_3^2)^2}\cong m_3^2~.
\end{equation}

A straightforward computation of the above quantities leads to the eigenvalues of the moduli, where we discriminate the eigenvalues of the third modulus into two cases. The first one corresponds to the limit of having fluxes of order one, while the second case describes the limit of exponentially small fluxes. This categorization will be needed for the subsequent analysis of the dark radiation predictions.

\begin{align}
Tr[M^2]\cong &\frac{72 d}{\mathcal{V}^2},\quad \begin{cases}
\dfrac{Det[M^2]}{Tr[M^2]^2}\cong -\frac{27 d^{1/3} \left(\frac{2}{3}q+1\right) }{8 q \mathcal{V}^{7/3}},\quad \mathcal{W}_0\sim \mathcal{O}(1),\\ 
\dfrac{Det[M^2]}{Tr[M^2]^2}\cong -\frac{27 d^{1/3} \mathcal{W}_0^{4/3} \left(\frac{2}{3}q+1\right) }{8 q \mathcal{V}^{7/3}},\quad \mathcal{W}_0\ll \mathcal{O}(1)~.
\end{cases}\label{masses}
\end{align}

The quantum correction $q$ acts as a key factor in the mass eigenvalues, where its scale along with the scale of the fluxes $\mathcal{W}_0$ will modify the masses in the spectrum. It would also be important to derive some useful bounds for the parameters of the theory using the above masses. Demanding the moduli $\tau_1,
\tau_2$ to be heavier compared to the transverse direction, which serves as the volume direction, a useful lower bound for the fluxes and the uplift parameter $d$ can be extracted. In Figure 2., the two cases are displayed for various values of the parameter $\eta$ along the allowed region of $q$, specified in \eqref{qpar}.

\begin{align}
    \dfrac{Tr[M^2]}{\frac{Det[M^2]}{Tr[M^2]^2}}=\begin{cases} 
    -\dfrac{64}{3} \dfrac{q d^{1/3}(\eta q \mathcal{W}_0^2)^{1/3}}{1+\frac{2q}{3}},\quad \mathcal{W}_0\sim \mathcal{O}(1)\\
    -\dfrac{64}{3} \dfrac{q d^{1/3}(\eta q)^{1/3}}{(1+\frac{2q}{3})(\mathcal{W}_0^2)^{1/3}},\quad \mathcal{W}_0\ll \mathcal{O}(1)~.
    \end{cases}
\end{align}

\begin{align}
    \dfrac{Tr[M^2]}{\frac{Det[M^2]}{Tr[M^2]^2}}>1 \Rightarrow \begin{cases}
        d \mathcal{W}_0^2> (-\dfrac{3(1+\frac{2q}{3})}{64 q(\eta q)^{1/3}})^3,\quad \mathcal{W}_0\sim \mathcal{O}(1)\\
        \frac{d} {\mathcal{W}_0^2}> (-\dfrac{3(1+\frac{2q}{3})}{64 q(\eta q)^{1/3}})^3,\quad \mathcal{W}_0\ll \mathcal{O}(1)\label{casesmasses}
    \end{cases}
\end{align}

\begin{figure}[H]
    \centering
    \includegraphics[scale=0.7]{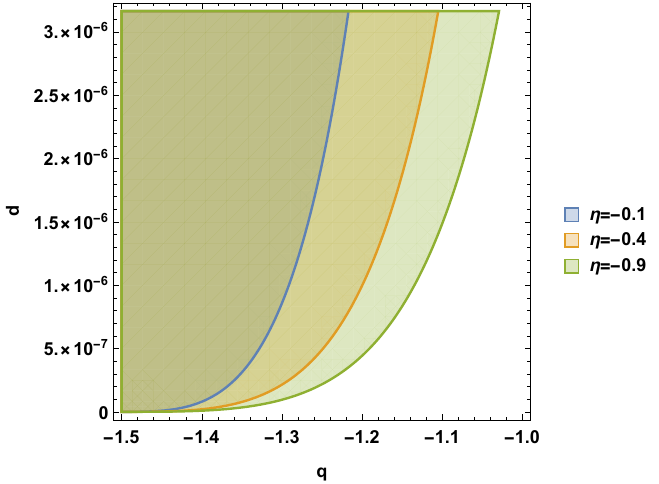}
    \quad
    \includegraphics[scale=0.7]{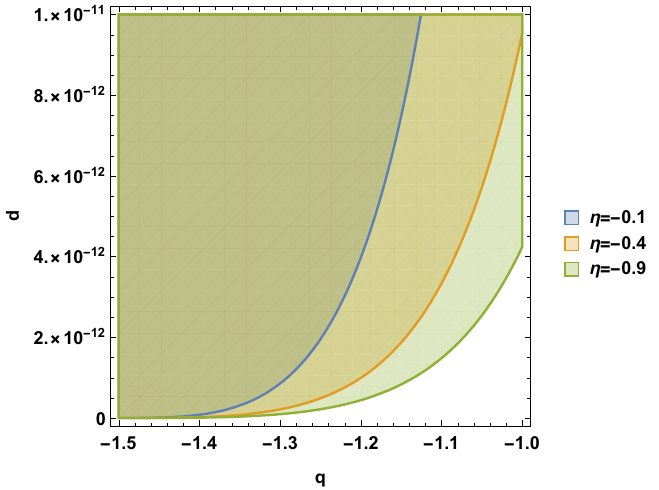}
    \caption{Left: Plot of the allowed $d$ values for the first case of \eqref{casesmasses}. The fluxes are set at $|\mathcal{W}_0|= 1$. Right: Plot of the allowed $d$ values for the second case of \eqref{casesmasses}. The fluxes are set at $|\mathcal{W}_0|= 10^{-3}$. }\label{dW0}
\end{figure}

As a next step, we proceed with the canonical normalization considering the computation of the eigenvectors of the mass matrix. In Appendix A.1, we explicitly compute the eigenvectors for the two cases describing the scale of the fluxes, where for both cases studies have shown that a dS vacuum is derived \cite{2203.03362, 1912.10047,2108.04266}. Beginning with the case $A)$ (exponentially suppressed fluxes $\mathcal{W}_0\ll 1$), the mass hierarchy and the mixing for the normalized fields are given by:

 \begin{equation}
        m^2_{\phi_2}\cong\dfrac{Det[M^2]}{Tr[M^2]^2}\ll m^2_{\phi_1} \cong  m^2_{\phi_3}\cong Tr[M^2]~.\label{case1}
\end{equation}

\begin{align}
    \tau_1 \cong  P_{12}^{A)} (\phi_2 +\phi_3) , \quad \tau_2 \cong   P_{21}^{A)} \phi_1 +  P_{23}^{A)} \phi_3,\quad \tau_3 \cong P_{32}^{A)} \phi_2 ~.\quad P_{32}^{A)}\gg P_{12}^{A)}~.
\end{align}

It is evident that the third modulus mainly parametrizes the $\phi_2$ field, while the $\tau_1$ modulus provides a small contribution to this field. On the other hand, the other directions are mostly described by the heavy moduli $\tau_1,\tau_2$ providing a clear geometric separation between the light volume direction and the perpendicular trajectories of the dS vacuum. In a similar fashion, for the case $B)$ (fluxes of order one $\mathcal{W}_0 \sim 1$, the same pattern of mixing between the moduli and the normalized fields is observed. The $\phi_2$ field is described by the lightest modulus $\tau_3$, while the transverse directions $\tau_1,\tau_2$ characterize the other two normalized fields
\footnote{Regarding the ratios between the mixing parameters $P_{ij}^{\alpha, \beta)}$, follow the discussion in Appendix A.1.}.    
    
\begin{equation}
    m^2_{\phi_2}\cong\dfrac{Det[M^2]}{Tr[M^2]^2}\ll m^2_{\phi_1} \cong  m^2_{\phi_3}\cong Tr[M^2]~.
\end{equation}

\begin{align}
    \tau_1 \cong  P_{12}^{B)} \phi_2 , \quad  \tau_2 \cong   P_{21}^{B)} \phi_1 +  P_{23}^{B)} \phi_3,\quad \tau_3 \cong P_{32}^{B)} \phi_2 ~.\quad P_{32}^{B)}\gg P_{12}^{B)}~.
\end{align}
    
 However, inflation is not our main object of study in this work, but we retain our mission for a consistent string embedding in a future work. Interestingly, if we would like to compare our model with the literature, the geometry discussed in \cite{2208.01017} and the references therein shares some characteristics with the above analysis. In their volume form, the visible sector is completely decoupled by the rest moduli and they separate two cases regarding the identification of the inflaton field. In the K{\"a}hler inflation case, the inflaton is denoted by the heaviest field and the transverse mode specifies the dark radiation predictions. Now, in the fibre inflation scenario of \cite{2208.01017}, the inflaton is identified by the lightest particle, just like our case $B)$. The same modulus specifies the reheating scale as well as the dark radiation predictions. Since the third direction of our setup is considerably lighter and more flat than the other two (as depicted in Figure 1.), the inflaton field could be identified with the $\phi_2$ field, i.e. the modulus $\tau_3$. This modulus mainly specifies the volume direction.

\section{Reheating and dark radiation predictions}

After the end of inflation, the inflaton will begin to oscillate around its minima, acquiring a large energy density in the process. Now, this energy density has to be transferred through a specific mechanism to the other fields of the theory, either to the visible sector or to the dark sector. Among this plethora of fields, there is a possibility that the late-time cosmological dynamics of the universe will be addressed not by the inflaton, but by a different field whose decay rate is much smaller. Thus, the final reheating temperature and the effective number of neutrino species will be determined by the decay rate of the aforementioned longest-lived particle. Considering the decay products of the moduli, they fall into two categories. The first are the decays that produce the particles of the visible sector, where these particles could be identified as either Higgs bosons or gauge bosons. We are going to employ a Giudice-Masiero mechanism to describe the relevant dynamics of the decays of K{\"a}hler moduli to the visible sector. In addition, there may also be decays to hidden sector states, which is a generic feature shared by string compactifications. The hidden sector contains several candidates for dark radiation, such as the axionic partner of the K{\"a}hler moduli fields or light hidden gauge bosons. Based on the analysis of the previous sections, we will identify the longest-lived particle in each case study with respect to the scale of the fluxes ($A)$ and $B)$), by explicitly computing the couplings of the moduli fields to the Higgs field and to the axions. Despite the criticism on these types of stringy constructions regarding the complex dynamics and the uncertainty with respect to the effective theory approximation, interesting proposals point toward the direction, where cosmological solutions could play a bilateral role. Firstly, these solutions provide a stringy origin for the reheating mechanism but also contribute to the identification of various dark matter particles \cite{2010.03573,2203.08833,1401.4364,1403.6473}.

We are going to start from the Lagrangian's kinetic terms, which can be expanded as:

\begin{align}
\mathcal{L}&= \mathcal{K}_{11}\partial_{\mu}\tau_1\partial^{\mu} \tau_1+\mathcal{K}_{12}\partial_{\mu}\tau_1\partial^{\mu} \tau_2+ \mathcal{K}_{13}\partial_{\mu}\tau_1\partial^{\mu} \tau_3+ \mathcal{K}_{21}\partial_{\mu}\tau_2\partial^{\mu} \tau_1+ \mathcal{K}_{22}\partial_{\mu}\tau_2\partial^{\mu} \tau_2\notag \\
&+\mathcal{K}_{23}\partial_{\mu}\tau_2\partial^{\mu} \tau_{3} + \mathcal{K}_{31}\partial_{\mu}\tau_3\partial^{\mu} \tau_1 +\mathcal{K}_{32}\partial_{\mu}\tau_3\partial^{\mu} \tau_2 + \mathcal{K}_{33}\partial_{\mu}\tau_3 \partial^{\mu} \tau_3+ V + \mathcal{O}(\dfrac{
\partial^2 V}{\partial_{\tau_i}\partial_{\tau_j}})\tau_i \tau_j~.
\end{align}

Since we aim to highlight the effect of the quantum corrected kinetic terms, we would like to include in the above Lagrangian, terms that contain cubic order interactions. More specifically, these terms are the interactions between the moduli and their corresponding axionic partners $c_i$. These trilinear interaction terms have the following structure:

\begin{align}
\mathcal{L}=(\partial_ {\tau_i} \mathcal{K}_{jk})\; \tau_i \partial_{\mu} c_j \partial^{\mu} c_k= (\partial_ {\tau_i} \mathcal{K}_{jk}) \frac{1}{2}(m^2_i -m^2_j-m^2_k) \tau_i c_j c_k,
\end{align}
where in the last step we used the Dirac equation, after integrated by parts, to recast the terms into their equivalent form containing their respective masses. As noted in \cite{2203.08833}, the cubic terms obtained by the derivatives of the potential are subleading compared to the ones originating from the kinetic terms. This fact can be attributed to the suppression due to the large volume expansion, where we expect a similar behavior in our geometry. We can assume that the masses of the axions are negligible compared to moduli's masses, concluding that only $m_i$ will contribute in the above interaction term. The dominant contribution in the above computation is the derivative of the K{\"a}hler metric with respect to each modulus of the parameter space, where the corresponding matrices are shown below. 

\begin{align}
\partial_ {\tau_1}\mathcal{K}_{ij}=\begin{pmatrix}
 -\frac{1}{2 d^2 \mathcal{V}^2} & \frac{\eta (6-q)}{4 d \mathcal{V}^3} & \frac{\eta (6-q)}{4 d \mathcal{V}^3} \\
\frac{\eta (6-q)}{4 d \mathcal{V}^3} & 0 & -\frac{\eta (q-5)}{12  \mathcal{V}^3} \\
 \frac{\eta (6-q)}{4 d \mathcal{V}^3} & -\frac{\eta (q-5)}{12  \mathcal{V}^3} & 0 \\
\end{pmatrix}~,
\end{align}

\begin{align}
\partial_ {\tau_2} \mathcal{K}_{ij}=\begin{pmatrix}
 0 & \frac{\eta (6-q)}{4  \mathcal{V}^3} & -\frac{\eta (q-5)}{12  \mathcal{V}^3} \\
\frac{\eta (6-q)}{4  \mathcal{V}^3} & -\frac{d}{2\mathcal{V}^2} &-\frac{d\eta (q-6)}{4  \mathcal{V}^3} \\
 -\frac{\eta (q-5)}{12  \mathcal{V}^3} & -\frac{d\eta (q-6)}{4  \mathcal{V}^3} & 0 \\
\end{pmatrix}~,\quad 
\partial_ {\tau_3}\mathcal{K}_{ij}=\begin{pmatrix}
 0 & \frac{\eta (-q+5)}{12  \mathcal{V}^{3}} & \frac{\eta (-q+6)}{4  \mathcal{V}^{3}} \\
\frac{\eta (-q+5)}{12  \mathcal{V}^{3}} & 0 & -\frac{d\eta(q-6)}{4\mathcal{V}^{3}}\\
 \frac{\eta (-q+6)}{4\mathcal{V}^{3}} & -\frac{d\eta(q-6)}{4\mathcal{V}^{3}} & -\frac{d}{2\mathcal{V}^{2}} \\
\end{pmatrix}~.\label{Kahlerder}
\end{align}

In the above matrices, the minimal values of the moduli have been applied, where we exchanged the $\tau_3$ modulus with the volume. Every coupling will be expressed only in terms of the free parameters of the theory, so it will be necessary to know the scaling with respect to the volume. Using all of the above ingredients, we could compute all the relevant coupling constants needed for the various decay rates. In doing so, the formulas in Appendix A.2 will be used. Before proceeding further, it is crucial to consider which is the longest-lived particle on this model, since this will determine the energy density of the universe. The most dominant contribution to the visible sector energy will come from the decay of the normalized field to Higgses. We focus on the Giudice-Masiero mechanism, where for the MSSM the relevant fields are the Higgses $H_u$ and $H_d$. Starting from an extended K{\"a}hler potential with the Higgses fields included, we will derive the formula of the decay rate of the process.

\begin{align}
    \mathcal{K} = -3 \ln \big[ (T_i+\bar{T}_i) + \dfrac{1}{3}(H_u \bar{H}_u + H_d \bar{H}_d + Z H_u \bar{H}_d)\big],
\end{align}

where the last term will be important for the subsequent discussion \cite{1208.3562,1208.3563,0805.2943}. The parameter Z in the Guidice-Masiero coupling is an undetermined constant (to leading order in
an inverse volume expansion). The relevant decay rate for the various moduli fields is given by:

\begin{align}
    \Gamma_{\tau_i \rightarrow H\bar{H}} \sim \dfrac{Z^2}{4\pi} \dfrac{m_{\phi}^3}{M_p^2}~.
\end{align}

Returning to our initial question, which concerns the determination of the longest-lived particle, it is important to recast the moduli fields $\tau_i$ to the normalized ones $\phi_i$, where this transition will contribute an additional factor (mixing factor) in the decay rate. This factor can be obtained from Appendix A.1. This is the point where the analysis of the previous section becomes useful. \par
Let us start from the case $A)$ of having exponentially suppressed fluxes, and we should compute the relevant decay rates of every normalized field to Higgses:

\begin{align}
\Gamma_{\phi_{2,3}(\tau_1)\rightarrow H\bar{H}}= \frac{432 \;Z^2 \sqrt{2} d^{29/6} (2 q+1)^2}{\pi (2 q+3)^2 \mathcal{V}^{7/3} \mathcal{W}_0^4}&,\quad 
\Gamma_{\phi_{1,3}(\tau_2)\rightarrow H\bar{H}}= -\frac{6912 \;Z^2\sqrt{2} d^{5/2}}{\pi \mathcal{V}^2 \mathcal{W}_0^2 (55 \eta -6 \xi )},  \\\notag \\
\Gamma_{\phi_2(\tau_3)\rightarrow H\bar{H}}=&\frac{27 \;Z^2\mathcal{W}_0^3 (-\eta  (2 q+3))^{3/2}}{\sqrt{2} \pi \mathcal{V}^4 (55 \eta -6 \xi )}~. 
\end{align}

where the modulus in the parentheses describes the modulus from which the normalized field originates. Given the decay rates above, the longest-lived particle can be characterized after comparing the ratios.

\begin{align}
   \dfrac{\Gamma_{\phi_{1,3}\rightarrow H\bar{H}}}{\Gamma_{\phi_2(\tau_3)\rightarrow H\bar{H}}}\cong -\frac{512 d^{5/2} \mathcal{V}^2}{\mathcal{W}_0^5 (-\eta  (2 q+3))^{3/2}}>1.
\end{align}

\begin{equation}
    \Gamma_{\phi_{1,3}\rightarrow H\bar{H}}>\Gamma_{\phi_2(\tau_3)\rightarrow H\bar{H}}~.
\end{equation}

Based on the ratios above between the decay rates, we observe that the smallest one is represented by $\Gamma_{\phi_2}$, making $\phi_2$ the longest-lived particle. This conclusion has important consequences for late-time cosmology, since the energy density will be determined by $\phi_2$. So, sketching up our geometric setup, we have to place the visible sector on the stack of branes represented by the $\tau_2$ world-volume, making this way the $\tau_1,\tau_3$ spaces the dark sectors of the geometry. Additionally, as pointed out in \cite{0711.3389}, any non-perturbative corrections along the cycles supporting the visible sector can not be allowed, since they will intersect with the chiral matter. Rephrasing the above argument, this type of construction could only allow axionic degrees of freedom from the transverse space of $\tau_2$. We have to make this particular choice, taking into account the fact that the lightest and longest-lived particle could in principle not only solve the dark radiation problem, but also their decays could produce the correct abundance of non-thermal dark matter. Consequently, $\phi_2$ (the $\tau_3$ modulus) cannot be identified with the visible sector.

For the case $B)$ of having order one fluxes, a similar computation could result in the following results for the decay rates of the normalized fields to Higgses.

\begin{align}
\Gamma_{\phi_2(\tau_1)\rightarrow H\bar{H}}= \frac{675\; Z^2 \sqrt{2} d^{29/6} \delta ^2}{\pi  \gamma ^2 \mathcal{V}^3}&,\;\;
\Gamma_{\phi_{1,3}\rightarrow H\bar{H}}= \frac{110592\;Z^2  \sqrt{2} \beta ^2 d^{3/2}}{25 \pi  \mathcal{V}^3 \mathcal{W}_0^2}, \\ \notag \\
\Gamma_{\phi_2(\tau_3)\rightarrow H\bar{H}}=-&\frac{27\;Z^2 d^{1/3} \delta ^2\; \mathcal{W}_0^3 (-\eta  (3+2q))^{3/2}}{16 \sqrt{2} \pi \gamma ^2 \mathcal{V}^5}, 
\end{align}

where the variables $\beta, \gamma, \delta$ are defined in Appendix A.1. Now, we could compare the respective decay rates to determine which one of the normalized fields can be considered in the given setup as the longest-lived particle.

\begin{align}
    \dfrac{\Gamma_{\phi_{1,3}\rightarrow H\bar{H}}}{\Gamma_{\phi_2(\tau_3)\rightarrow H\bar{H}}}&=-\frac{131072 d^{1/6} \left(\eta  q \mathcal{W}_0^2\right){}^{7/3} \left(-8 (d \eta  q \mathcal{W}_0^2)^{1/3}+2 q+3\right){}^2}{25 (2 q+1)^2 \mathcal{W}_0^5 (6 \xi-55 \eta ) (-\eta  (3+2q))^{3/2}}, \\ \notag \\
    &\dfrac{\Gamma_{\phi_{1,3}\rightarrow H\bar{H}}}{\Gamma_{\phi_2(\tau_3)\rightarrow H\bar{H}}}\cong  \mathcal{O}(\dfrac{10^4d(3+2q)^{1/2}}{(-\eta)^{3/2}})>1.
\end{align}

\begin{equation}
    \Gamma_{\phi_{1,3}\rightarrow H\bar{H}}>\Gamma_{\phi_2(\tau_3)\rightarrow H\bar{H}}~.
\end{equation}

From the above, we can conclude that the normalized field $\phi_2$ corresponds to the longest-lived particle in this geometric setup. Again, as previously, we need to specify the visible and hidden sectors of the internal geometry. In this case, the visible sector will be identified with the $\tau_2$ stack of branes, while the $\phi_2$ field will be important in both discussions of dark radiation and dark matter.

Returning to our main purpose, which is to calculate the dark radiation predictions of this model, we need to identify the relevant decay channels. The decay rates of the longest-lived particle to axions must be examined, and we first write the interaction terms associated to these processes.

\begin{align}
    \mathcal{L} \supset (\partial_ {\tau_m} \mathcal{K}_{np})\; \tau_m \partial_{\mu} c_n \partial^{\mu} c_m =\partial_{\tau_m} \mathcal{K}_{np} P^{\tau}_{mi} P^{c}_{nj} P^{c}_{pk} \phi_i \partial_{\mu}\alpha_j \partial^{\mu}\alpha_k= \mathcal{K}_{mnp} P^{\tau}_{mi} P^{c}_{nj} P^{c}_{pk} m^2_{\phi_i} \phi_i \alpha_j \alpha_k,\label{coupling}
\end{align}

where in the last step we also transformed the axionic partner $c_i$ of the K{\"a}hler modulus to the normalized axion $\alpha_i$. The $P_{ij}$ represent elements of the mixing matrix, when we apply the basis transformation. The mixing between the axions will be determined by studying the induced scalar potential, when non-perturbative corrections are included in the superpotential $\mathcal{W}$. Geometric constructions, which include both perturbative and non-perturbative corrections, have been studied in the past \cite{2007.15423, 2109.08421} and the problem of moduli stabilization remains unaffected. The large volume limit is necessary to ensure the validity of the effective theory and to keep the exponential factor of the non-perturbative corrections $\mathcal{W} \supset A e^{-a \tau_i}$ parametrically small. Since we focus on the impact of quantum corrections, we can parametrize this matrix as follows and focus on the qualitative behavior of the decay rates \footnote{For an extended discussion in a string derived model, follow Appendix B in \cite{2203.08833}.}.

\begin{align}
    \begin{pmatrix}
        c_1\\
        c_2\\
        c_3
    \end{pmatrix}= \begin{pmatrix}
        \lambda_{11} & \lambda_{12} &\lambda_{13}\\
        \lambda_{21} & \lambda_{22} & \lambda_{23}\\
         \lambda_{31} & \lambda_{32} & \lambda_{33}\\
    \end{pmatrix}\begin{pmatrix}
        \alpha_1\\
        \alpha_2\\
        \alpha_3
    \end{pmatrix}.\label{axmix}
\end{align}

For our first case study $A)$, the longest-lived particle is $\phi_2$, which descends from the $\tau_3$ modulus. Observing the Lagrangian terms in equation \eqref{coupling}, the $m,i$ indices are determined from our previous analysis, so $m=3,i=2$. For the $j,k$ indices, these can only span $(j,k)=(1,3)$, since the visible sector is represented by the $\tau_2$ modulus. The relevant Lagrangian part for this process can be written, after summing all the contributions, as:

\begin{align}
     \mathcal{L} \supset  \mathcal{K}_{3np} P^{\tau}_{32} P^{c}_{nj} P^{c}_{pk} m^2_{\phi_2} \phi_2 \alpha_j \alpha_k~,
\end{align}

where we inserted $i=2$ for the $\phi_2$ normalized field and $m=3$, since this field descends from the $\tau_3$ modulus. The coupling constant for the decay of $\phi_2$ to the $\alpha_3$ axion (a similar analysis can be applied for the $\alpha_1$ axion) is given by the following:

\begin{align}
    \phi_2 \rightarrow \alpha_3 \alpha_3 : \; \mathcal{K}_{3np}P^{\tau}_{32} P^c_{n3} P^c_{p3} m^2_{\phi_2}\phi_3 \alpha_3 \alpha_3 = P^{\tau}_{32}(\mathcal{K}_{333} P^c_{33} P^c_{33} + 2\mathcal{K}_{313} P^c_{13} P^c_{33} + 2\mathcal{K}_{312} P^c_{13} P^c_{23} )m^2_{\phi_2}\phi_2 \alpha_3 \alpha_3
\end{align}

\begin{align}
    P^{\tau}_{32}(\mathcal{K}_{333} P^c_{33} P^c_{33})m^2_{\phi_2}= &\frac{9 \beta  \eta  \lambda _{33}^2 (2 q+3) \mathcal{W}_0^3 (\eta  q)^{1/3}}{2 \mathcal{V}^{17/3}}, \quad P^{\tau}_{32}(2\mathcal{K}_{313} P^c_{13} P^c_{33})m^2_{\phi_2}=\frac{9 \beta  \eta ^2 \lambda _{13} \lambda _{33} (q-6) (2 q+3) \mathcal{W}_0}{2 \mathcal{V}^{17/3} (\eta  q)^{2/3}},\notag \\
    &P^{\tau}_{32}(2\mathcal{K}_{312} P^c_{13} P^c_{23})m^2_{\phi_2}=\frac{3 \beta  \eta ^2 \lambda _{13} \lambda _{23} (q-5) (2 q+3) \mathcal{W}_0}{2 \mathcal{V}^{17/3} (\eta  q)^{2/3}}~.
\end{align}

\begin{align}
    \dfrac{P^{\tau}_{32}(\mathcal{K}_{333} P^c_{33} P^c_{33})m^2_{\phi_2}}{P^{\tau}_{32}(2\mathcal{K}_{313} P^c_{13} P^c_{33})m^2_{\phi_2}} = \dfrac{q \mathcal{W}_0^2}{-6+q} \dfrac{\lambda_{33}}{\lambda_{13}}, \quad 
    \dfrac{P^{\tau}_{32}(\mathcal{K}_{333} P^c_{33} P^c_{33})m^2_{\phi_2}}{P^{\tau}_{32}(2\mathcal{K}_{312} P^c_{13} P^c_{23})m^2_{\phi_2}} = \dfrac{3q \mathcal{W}_0^2}{q-5} \dfrac{\lambda_{33}^2}{\lambda_{13}\lambda_{23}}~.
\end{align}

One can observe that previous studies had omitted the contributions of the off-diagonal entries in the K{\"a}hler metric. It is also worth mentioning that the contribution from $\mathcal{K}_{313}$ is more dominant compared to $\mathcal{K}_{312}$. As for the relevant scale between the diagonal and the off-diagonal entries of matrix \eqref{axmix}, it is expected that $\lambda_{ii}>\lambda_{ij}$ since we assume a normal ordering in the mass hierarchy of the axions. Next, we compute the total decay rate to the axions as follows:

\begin{align}
    \Gamma_{tot} = \frac{3 \beta ^2 \eta ^3 \mathcal{W}_0^3 \left(-\frac{2 q+3}{\eta }\right)^{3/2} \left(3 d \lambda _{33}^2 \mathcal{V}+\eta  \lambda _{13} \left(\lambda _{23} (q-5)+3 \lambda
   _{33} (q-6)\right)\right){}^2}{8 \sqrt{2} \pi  d \mathcal{V}^{11}} M_p~.\label{axion1}
\end{align}

Based on the above, we are at a point where we need to connect our theoretical computations to the observable quantities such as the effective number of neutrino species and the reheating. The definition of the effective number of neutrino species is given by: 

\begin{align}
    \Delta N_{eff}=\dfrac{43}{7}(\dfrac{10.75}{g_*(T_{rh})})^{1/3} \dfrac{\Gamma_{tot}}{\Gamma_{\phi_1\rightarrow H \bar{H}}}= \dfrac{43}{7}(\dfrac{10.75}{g_*(T_{rh})})^{1/3} \frac{ \left(3 d \lambda _{33}^2 \mathcal{V}+\eta  \lambda _{13} \left(\lambda _{23} (q-5)+3 \lambda _{33} (q-6)\right)\right){}^2}{72 \mathcal{V}^6} ,
\end{align}

where $g_*(T_{rh})$ denotes the number of relativistic degrees of freedom at the reheating temperature $T_{rh}$. Additionally, we are going to define a new quantity for the complete decay rate (both to visible and dark sectors):

\begin{equation}
    \Gamma^{\prime} = \Gamma_{\phi_1 \rightarrow H \bar{H}}+ \Gamma_{tot}~.
\end{equation}

Using this quantity, we can straightforwardly compute the reheating temperature of this model. This is given by:

\begin{align}
    T_{rh} = (\dfrac{90}{\pi g_*(T_{rh})}\dfrac{\Gamma_{\phi_1 \rightarrow H \bar{H}}}{\Gamma^{\prime}})^{1/4}\sqrt{\Gamma^{\prime} M_p}~.
\end{align}

Since the decays of the moduli to axions are highly suppressed compared to the ones on Higgses, the complete decay rate can be effectively described by the decay rate to the visible sector. The reheating temperature is then described by the following:

\begin{align}
     T_{rh} \cong (\dfrac{90}{\pi g_*(T_{rh})})^{1/4}\sqrt{\Gamma_{\phi_1 \rightarrow H \bar{H}} M_p} =  (\dfrac{90}{\pi g_*(T_{rh})})^{1/4} \frac{3 \sqrt{3}  (-\eta  (2 q+3))^{3/4}}{2^{1/4} \pi^{1/2}  \mathcal{V}^2 \sqrt{(55 \eta -6 \xi )\mathcal{W}_0^{-3}}}Z\; M_p~.\label{reh1}
\end{align}

The cosmological moduli problem \cite{Coughlan:1983ci,hep-ph/9308292}, also, constrains the number of fluxes in the present model. The lightest modulus is bounded from below to lay at scales $\mathcal{O}(10) \text{TeV}$, resulting in:

\begin{equation}
    m_{\phi_2} > 10 \; \text{TeV} \Rightarrow  \mathcal{W}_0^2>-\frac{800 \mathcal{V}^{8/3} (\eta  q)^{2/3}}{27 \eta   \left(\frac{2 q}{3}+1\right)} \frac{\text{TeV}^2}{M_p^2}~.
\end{equation}

Given this very restrictive bound, we would like to comment on that regarding the geometry of the compactified space. Our case study suggests that we are exploring exponentially suppressed integer fluxes. The toroidal structure of the volume form cannot admit a solution for arbitrary small fluxes. Moreover, we expect that our results will not change significantly by considering higher order terms in the effective potential, since they are subleading due to the suppression because of $\eta^n$ and $\frac{\xi}{\mathcal{V}^n}$. This fact could also provide a bottom-up proof of why this geometry, previously studied in \cite{1803.08941}, accommodates more easily stabilized solutions with order-one fluxes. In addition, only moderate values of volume are accepted in this case, since the masses of the moduli would be below the bound presented above. It is imperative to mention that these cosmological implications of stringy constructions could also be used as a testing tool to clarify the properties of the background geometry and the scale of the fluxes $\mathcal{W}_0$ \cite{1912.10047, 2107.09064}. Moreover, we illustrate three numerical examples (Table 1.) for various scales of reheating temperature of our case study $A)$. In addition, the moduli masses at the stable vacuum are presented in Table 2.

\begin{table}[H]
\begin{center}
\small
\begin{tabular}{|c||c|c|c|c|c|c|c|c|c|}
\hline
  & $|\mathcal{W}_0|$ & $\eta$ & $\xi$ & $d$ & $T_{rh}$\; (\text{MeV})  & $g_*(T_{rh})$ & $Z$ & $q$  \\
\hline
 $\mathcal{V}\cong$ 2859 & $1 \times 10^{-9}$ & -0.9 & 5 & $4\times 10^{-22}$ & $8.5$ & $10.75$ & 1 &$-1.27$\\
 \hline
 $\mathcal{V}\cong$ 4314 & $1\times 10^{-10}$ & -0.9 & 6 & $2.2\times 10^{-24}$  & $4.5$ & $10.75$ & 20 & $-1.05$\\
 \hline
\end{tabular}
\end{center}
\caption{Different reheating temperatures for various set of parameters. Obviously, exponentially small fluxes tend to reproduce a low reheating scenario. The effective neutrino number of species tends to its Standard model value $\Delta N_{eff}\rightarrow 0$.}\label{Case1}
\end{table}%

\begin{table}[H]
\begin{center}
\small
\begin{tabular}{|c||c|c|c|}
\hline
 $m^2_{\phi_i} (\text{TeV})$ & $m^2_{\phi_1}$ & $m^2_{\phi_3}$ & $m^2_{\phi_2}$   \\
\hline
 $|\mathcal{W}_0 |= 1 \times 10^{-9}$  & $712$ & $712$ & $192$ \\
 \hline
 $|\mathcal{W}_0| = 1 \times 10^{-10}$  & $35$ & $35$ & $17$ \\
 \hline
\end{tabular}
\end{center}
\caption{The moduli masses along the two numerical examples presented above. }\label{Case1}
\end{table}%

Following the above discussion, we discuss the scenario of fluxes of order $\mathcal{W}_0\sim \mathcal{O}(1)$. The longest-lived particle in this case (case $B)$) is $\phi_2$. The relevant Lagrangian part is written as:

\begin{align}
     \mathcal{L} \supset  \mathcal{K}_{3np} P^{\tau}_{mi} P^{c}_{nj} P^{c}_{pk} m^2_{\phi_2} \phi_2 \alpha_j \alpha_k~,
\end{align}

where we inserted $i=3$ for the $\phi_3$ normalized field and $m=2$, since this field descends from the $\tau_3$ modulus. The first coupling constant is computed:

\begin{align}
    \phi_2 \rightarrow \alpha_3 \alpha_3 : \; \mathcal{K}_{3np}P^{\tau}_{32} P^c_{n3} P^c_{p3} m^2_{\phi_2}\phi_3 \alpha_3 \alpha_3 = P^{\tau}_{32}(\mathcal{K}_{333} P^c_{33} P^c_{33} + 2\mathcal{K}_{313} P^c_{13} P^c_{33} + 2\mathcal{K}_{312} P^c_{13} P^c_{23} )m^2_{\phi_2}\phi_2 \alpha_3 \alpha_3
\end{align}

\begin{align}
    P^{\tau}_{32}(\mathcal{K}_{333} P^c_{33} P^c_{33})m^2_{\phi_2}= &\frac{9 d \delta  \eta  \lambda _{33}^2 (2 q+3) \mathcal{W}_0^2}{8 \gamma  \mathcal{V}^{16/3}}, \quad P^{\tau}_{32}(2\mathcal{K}_{313} P^c_{13} P^c_{33})m^2_{\phi_2}=\frac{9 \delta  \eta ^2 \lambda _{13} \lambda _{33} (q-6) (2 q+3) \mathcal{W}_0^2}{8 \gamma  \mathcal{V}^{19/3}},\notag \\
    &P^{\tau}_{32}(2\mathcal{K}_{312} P^c_{13} P^c_{23})m^2_{\phi_2}=\frac{3 \delta  \eta ^2 \lambda _{13} \lambda _{23} (q-5) (2 q+3) \mathcal{W}_0^2}{8 \gamma  \mathcal{V}^{19/3}}~.
\end{align}

\begin{align}
    \dfrac{P^{\tau}_{32}(\mathcal{K}_{333} P^c_{33} P^c_{33})m^2_{\phi_2}}{P^{\tau}_{32}(2\mathcal{K}_{313} P^c_{13} P^c_{33})m^2_{\phi_2}} = \dfrac{q \mathcal{W}_0^2}{-6+q} \dfrac{\lambda_{33}}{\lambda_{13}}, \quad 
    \dfrac{P^{\tau}_{32}(\mathcal{K}_{333} P^c_{33} P^c_{33})m^2_{\phi_2}}{P^{\tau}_{32}(2\mathcal{K}_{312} P^c_{13} P^c_{23})m^2_{\phi_2}} = \dfrac{3q \mathcal{W}_0^2}{q-5} \dfrac{\lambda_{33}^2}{\lambda_{13}\lambda_{23}}~.
\end{align}

It is evident, again, as to the former case, that the off-diagonal entries give rise to new contributions that were previously underestimated, as previously stated. Based on the above, we have to sum all the contributions, where this results to:

\begin{align}
    \Gamma_{tot} = -\frac{3 d^{1/3}\; \delta ^2 \mathcal{W}_0^3 (-\eta  (2 q+3))^{3/2} \left(3 d \lambda _{33}^2 \mathcal{V}+\eta  \lambda _{13} \left(\lambda _{23} (q-5)+3 \lambda _{33}
   (q-6)\right)\right){}^2}{128 \sqrt{2} \pi  \gamma ^2 \mathcal{V}^{11}} M_p~.
\end{align}

In this case, the standard definition of the effective number of neutrino species is given by: 

\begin{align}
    \Delta N_{eff}=\dfrac{43}{7}(\dfrac{10.75}{g_*(T_{rh})})^{1/3} \dfrac{\Gamma_{\phi_3 \rightarrow \alpha_3 \alpha_3}}{\Gamma_{\phi_3\rightarrow H \bar{H}}}= \dfrac{43}{7}(\dfrac{10.75}{g_*(T_{rh})})^{1/3} \frac{  \left(3 d \lambda _{33}^2 \mathcal{V}+\eta  \lambda _{13} \left(\lambda _{23} (q-5)+3 \lambda _{33} (q-6)\right)\right){}^2}{72 Z^2\mathcal{V}^6},
\end{align}

Now, the reheating temperature is computed as:

\begin{align}
    T_{rh}\cong 
    (\dfrac{90}{\pi g_*(T_{rh})})^{1/4} \frac{3 \sqrt{3} d^{1/6}\;  (2 q+1) \left(-\mathcal{W}_0\right){}^{3/2} (-\eta  (2 q+3))^{3/4}}{2^{9/4}\; \pi^{1/2}  \gamma  \mathcal{V}^{13/6}} Z\;M_p~.\label{reh2}
\end{align}

Finally, we will present various numerical examples (Table 3. and Table 4.) for the case of having order one fluxes. This scenario results in high-scale reheating temperatures, while the effective number of neutrinos species remains highly suppressed even for order-one coupling constants. This means that $\Delta N_{eff} \rightarrow 0$ is very close to the value of Standard Model.

\begin{table}[H]
\begin{center}
\small
\begin{tabular}{|c||c|c|c|c|c|c|c|c|c|}
\hline
  & $|\mathcal{W}_0|$ & $\eta$ & $\xi$ & $d$ & $T_{rh}$\; (\text{GeV}) & $g_*(T_{rh})$& $Z$  \\
\hline
 $\mathcal{V}\cong$ 2859 & $1$ & -0.9 & 5 & $4\times 10^{-4}$ & $2.3\times 10^{11}$ & $106.75$  & 1\\
 \hline
 $\mathcal{V}\cong$ 4314 & $5$ & -0.9 & 6 & $5.5\times 10^{-3}$  & $9.8\times 10^{10}$ & $106.75$  & 1\\
 \hline
\end{tabular}
\end{center}
\caption{Different reheating temperatures for various set of parameters. Obviously, large fluxes tend to reproduce a high scale reheating scenario. The effective number of neutrino species $\Delta N_{eff} \rightarrow 0$ is well below the allowed upper bound even for order $\mathcal{O}(0.1)$ coupling $\lambda^{\prime \prime}$.}\label{Case1}
\end{table}%

\begin{table}[H]
\begin{center}
\small
\begin{tabular}{|c||c|c|c|}
\hline
 $m^2_{\phi_i} (\text{TeV})$ & $m^2_{\phi_1}$ & $m^2_{\phi_3}$ & $m^2_{\phi_2}$    \\
\hline
 $|\mathcal{W}_0| = 1 $  & $7\times 10^{14}$ & $7\times 10^{14}$ & $2\times 10^{14}$ \\
 \hline
 $|\mathcal{W}_0 |= 5$  & $1.8\times 10^{15}$ & $1.8\times 10^{15}$ & $2.8\times 10^{14}$ \\
 \hline
\end{tabular}
\end{center}
\caption{The moduli masses along the two numerical examples presented above. }\label{Case1}
\end{table}%

\section{Dark matter scenario in the presence of quantum effects}

In various string models, cosmological predictions tend to suggest non-thermal dark matter (DM), produced by decays of heavy scalars. The most common production process involves the decay of the lightest modulus that provides a large amount of entropy, diluting any previous DM abundance, and then its byproducts would yield the necessary relic density. Possible dark matter particles fall into two categories regarding their thermodynamic origin and more specifically the freeze-out temperature $T_f\sim m_{DM}/20$.
\begin{itemize}
    \item $(T_{rh}>T_f):$ In this case dark matter particles are in a thermodynamical equilibrium and the annihilations between themselves favor the thermal origin of dark matter.
    \item $(T_{rh}<T_f):$ Here, a discrimination with respect to the dark matter annihilations is needed to be stated: the efficiency of the aforementioned annihilations is quantified in the critical abundance $Y_{DM}^c$ of dark matter particles, which quantity is obtained by the Boltzmann equations.
    \begin{align}
        Y_{DM}^c\cong \dfrac{H}{\langle \sigma_{ann} v \rangle_s}|_{T_{rh}}~.
    \end{align}
\end{itemize}
In the first sub-case, there is the possibility that the produced abundance exceeds the critical value $Y_{DM}>Y^c_{DM}$. Consequently, there exists some further time for annihilations until the two quantities match, a scenario labeled as "Annihilation scenario". The final relic abundance in this case is computed as:
    \begin{align}
        Y^c_{DM}\sim (\dfrac{n_{DM}}{s})_{obs} \dfrac{\langle \sigma_{ann}v \rangle^{th}_f}{\langle \sigma_{ann}v \rangle_f} \dfrac{T_f}{T_{rh}},\label{annihilation}
    \end{align}
where $n_{DM}$ represents the number density and $\langle \sigma_{ann}v \rangle^{th}_f\sim 3\times 10^{-26} \; \text{cm}^3 s^{-1}$ stands for the cross section of the thermal case as a requirement to produce the observed value of dark matter abundance \cite{1611.03184}

    \begin{align}
        (\dfrac{n_{DM}}{s})_{obs} \cong 5\times 10^{-10} (\dfrac{\text{GeV}}{m_{DM}})~.\label{obs}
    \end{align}
    
Readily, processes of this type are enhanced by a factor of $\dfrac{T_f}{T_{rh}}$ as opposed to the thermal production, where possible dark matter particles could be thermally underproduced Higgsino-like or Wino-like particles. In the second sub-case, the dark matter abundance is lower compared to the critical value, making the annihilations more scarce. This scenario is denoted as the "Branching scenario", where the resulting abundance is simply given by:
    \begin{align}
    Y_{DM}=Y_{\phi}\text{Br}_{DM}, \quad Y_{\phi}=\dfrac{3T_{rh}}{4m_{\phi}},\label{dmbr}
    \end{align}

where $Y_{\phi}$ calculates the abundance of the normalized fields and $Br_{DM}$ quantifies the decay ratio. In this case, Bino-like particles are favored for lower values of $Br_{DM}$ \cite{2308.16242, 2308.15380, 2010.03573, 2312.00136, 2402.18557}.

Following the analysis of the previous section, we will use the computed reheating temperature to scan the viability of both scenarios in the presence of the newly introduced quantum effects. According to \cite{1401.4364}, the branching scenario would better describe the parametric region of low-scale reheating temperatures.

\begin{align}
    \mathcal{O}(MeV)< T_{rh} \leq \mathcal{O}(GeV)~.\label{BrCon}
\end{align}

We can easily compute the abundance of the normalized fields, using the reheating temperature as defined in \eqref{reh1}:

\begin{align}
    Y_{\phi}=-\frac{9\times 5^{1/4}\; \beta  q \;(-\eta ^3 (2 q+3))^{1/4}Z}{\sqrt{2} \pi ^{3/4}\; (g_*(T_{rh}))^{1/4}\; \sqrt{q \mathcal{W}_0} (\eta  q \mathcal{V})^{2/3}}~.
\end{align}

For consistency, we are going to use the values of Table 1. and various plots will be presented to show the bounds on the parameters and the dark matter mass. In the branching scenario, the dark matter abundance has to be equal to the observed value \eqref{obs}, where the dark matter mass is bounded from below as follows:

\begin{align}
     T_f > T_{rh} \Rightarrow m_{DM}> \frac{180Z \times(5)^{1/4}  \sqrt{-\frac{\beta ^2 M_p^2 \mathcal{W}_0 (-\eta  (2 q+3))^{3/2}}{\eta  q \mathcal{V}^4}}}{(g_*(T_{rh}))^{1/4}\pi ^{3/4}}~,\label{massbound}
\end{align}

\begin{align}
    -\mathcal{W}_0< \dfrac{\eta q \mathcal{V}^4}{\beta^2 (-\eta(3+2q))^{3/4}} \dfrac{\pi^{3/2}(g_*(T_{rh}))^{1/2}}{(180\times 5^{1/4})^2} \dfrac{m_{DM}^2}{ZM_p^2}~.
\end{align}

The above bound for the scale of the fluxes is derived by solving with respect to the fluxes the inequality $T_f > T_{rh}$. In Figure 3., we depict the dark matter mass required to satisfy the observed dark matter abundance. In addition, we would like to incorporate in the present study, an interesting scenario that could explain the Baryon-Dark Matter Coincidence. As suggested in \cite{1011.1286}, late-time decays of moduli could produce some N species with C- and CP- violating coupling with the degrees of freedom of the visible sector. To quantify the above statement, the baryon asymmetry of the universe is given by:

\begin{equation}
    \eta_B \equiv \dfrac{\eta_B - \eta_{\bar{B}}}{s} = Y_{\phi} Br_N \epsilon\cong 9\times 10^{-11},
\end{equation}

where $\epsilon$ is the generated asymmetry per N decay and $Br_N$ denotes the branching ratio of the decays to the particles in the visible sector. By counting the degrees of freedom of the visible sector, the branching ratio could be of order $Br_N\sim \mathcal{O}(10^{-2}-10^{-1})$ \cite{1005.2804, 1011.1286}. So, by considering the branching scenario in this model, dark matter production could also resolve the baryon asymmetry problem. The Baryon-Dark Matter Coincidence is given by the following formula.

\begin{equation}
    \dfrac{\Omega_B}{\Omega_{DM}} \cong \dfrac{\text{GeV}}{m_{DM}}\dfrac{\epsilon Br_N}{Br_{DM}}\cong \dfrac{1}{6}\Rightarrow \epsilon Br_N\cong \dfrac{5\times 10^{-10}}{6 Y_{\phi}}~.
\end{equation}

where in the last we used the definitions in \eqref{obs}, \eqref{dmbr}. From the following plots, we can deduce that the for low-scale reheating scenarios, a correct prediction for the dark matter abundance and the solution for the Baryon-Dark Matter Coincidence can be derived. The observed value for the baryon asymmetry is set at $9\times 10^{-11}$, which gives us the required values for the branching ratio $Br_N,$ and the generated asymmetry $\epsilon$ produced by the decays of the N species particles. Moreover, in Appendix A2. we provide an alternative scenario where the string loop corrections are suppressed ($\eta \sim 0.1$), which leads to an inconsistent scenario and the Baryon-Dark Matter Coincidence cannot be cannot be explained.   

\begin{equation}
    \eta_B\cong 9\times 10^{-11}\Rightarrow Br_N \epsilon \cong 2.3\times 10^{-3}~.
\end{equation}

\begin{figure}[H]
    \centering
    {{\includegraphics[scale=0.665]{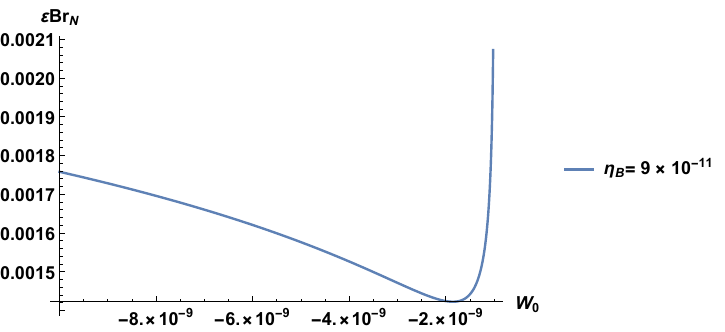} }}
    \quad
    {{\includegraphics[scale=0.665]{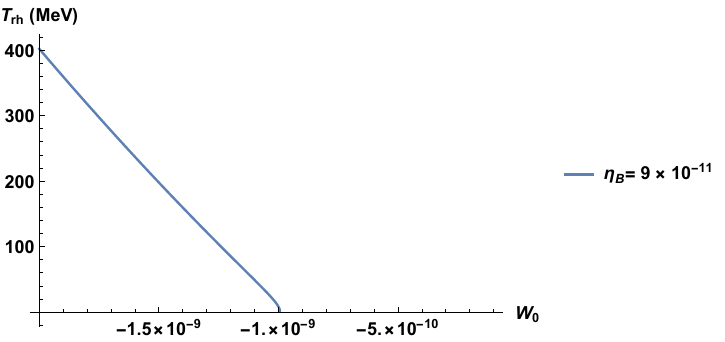} }}\\
    {{\includegraphics[scale=0.67]{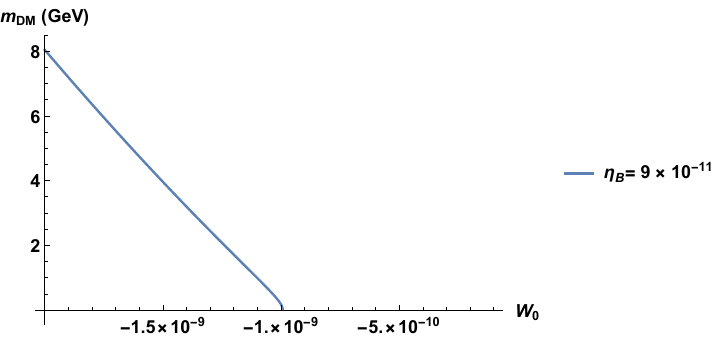} }}\quad
    {{\includegraphics[scale=0.67]{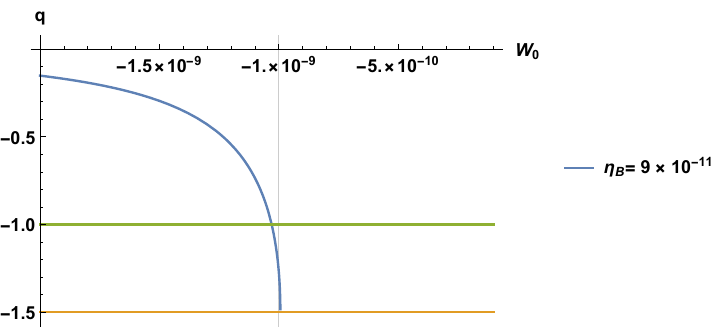} }}
    \caption{Plots for the first numerical example of Table 1. Upper left plot depicts the product of asymmetry with branching ratio with respect to the scale of fluxes, while the rest of the plots shows the required reheating temperature  and the scale of the dark matter mass. In the lower right plot, the yellow horizontal lines characterize the allowed values for the parameter $q$ in order to have dS vacuum for the effective potential.} 
\end{figure}

Regarding the annihilation scenario, it would fit better in the case of having high-scale reheating temperatures $T_{rh}\gg \text{GeV}$. So, the case of large integer fluxes could be embedded in this scenario. The reheating temperature, using the equation \eqref{reh2}, in this case is given by:

\begin{align}
    T_{rh}=\frac{9\;Z \times (5)^{1/4}  \sqrt{-\frac{d^{1/3}\; \delta ^2  \mathcal{W}_0^3 (-\eta  (2 q+3))^{3/2}}{\gamma ^2 \mathcal{V}^5}}}{4 (g_*(T_{rh}))^{1/4}\pi^{3/4}} M_p~.
\end{align}

From equation \eqref{annihilation}, we can see that the final relic abundance is given in terms of the number density and the freeze-out temperature $T_f$. A lower bound for the mass of the dark matter particle can be derived by requiring the freeze-out temperature to exceed the $T_{rh}$,

\begin{align}
    T_f > T_{rh} \Rightarrow m_{DM} >\frac{45\;Z \times (5)^{1/4}  \sqrt{-\frac{d^{1/3}\; \delta ^2  \mathcal{W}_0^3 (-\eta  (2 q+3))^{3/2}}{\gamma ^2 \mathcal{V}^5}}}{ (g_*(T_{rh}))^{1/4}\pi^{3/4}} M_p~.\label{boundan}
\end{align}

In Figure 4., the bound on the dark matter mass is depicted based on the example studied in Table 3. The horizontal lines represent the bound of the equation \eqref{boundan}, while the curve sketches the observed value of the dark matter mass at the limit $T_f \cong 20\; T_{rh}$. It is evident that the high-scale reheating scenario can saturate the dark matter abundance if the mass resides at high energies, rendering a superheavy candidate in such a paradigm.

\begin{align}
     Y^c_{DM}\sim (\dfrac{n_{DM}}{s})_{obs} \dfrac{\langle \sigma_{ann}v \rangle^{th}_f}{\langle \sigma_{ann}v \rangle_f} \dfrac{T_f}{T_{rh}}\Rightarrow T_{rh}= \dfrac{3\times 10^{-26} cm^3 s^{-1}}{\langle \sigma_{ann}v \rangle_f} \dfrac{m_{DM}}{20},\quad \kappa = \frac{\langle \sigma_{ann}v \rangle_f^{th}}{\langle \sigma_{ann}v \rangle_f}
\end{align}

\begin{figure}[H]
    \centering
    {{\includegraphics[scale=1]{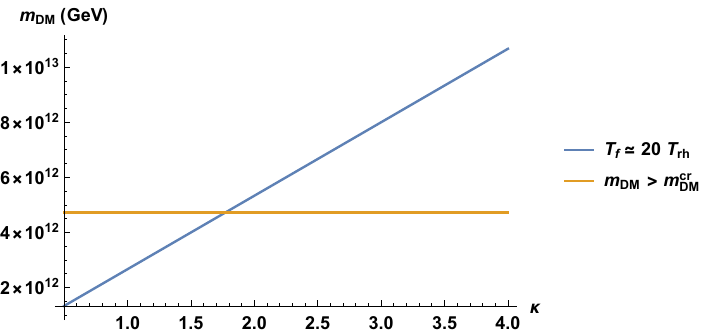} }}
    \caption{Dark matter mass with respect to the ratio of cross sections. The horizontal line represents the lower bound on the allowed mass. The parameters of the first numerical example of Table 3. are used for this plot.}
\end{figure}

Thus, we see that the annihilation scenario in the case of order one fluxes renders some superheavy dark matter particles with their mass being larger than $m_{DM}>10^{12} \; \text{GeV}$.

\section{Conclusions}

In this paper, we examined the impact of string loop corrections on various open problems of the early universe physics, such as the origin of dark radiation and dark matter. Firstly, we summarized their effects on the moduli stabilization procedure and provided a comprehensive explanation regarding the stability of dS vacua, placing very stringent bounds on the free parameters of the theory through $q=(d\mathcal{V}_{min})/(\eta \mathcal{W}_0^2)$. \par 

Subsequently, the K{\"a}hler metric was augmented by off-diagonal corrections, which depend on higher order corrections in terms of $\xi,
\;\eta$, and the moduli mass hierarchy was categorized with respect to the scale of the integer fluxes. Two different scenarios were investigated, the first one with exponentially suppressed fluxes and the latter with order one fluxes, where the moduli mixing with the canonical normalized fields provide a geometric separation of the different sectors of the theory. Furthermore, the Giudice-Masiero mechanism was incorporated to the model to study the decays of the moduli to the Standard model degrees of freedom, where it is shown that the longest-lived particle is associated with the lightest modulus which couples subdominantly with the other fields. Having this information, we provided a detailed analysis of the decays rates of the moduli to axions, taking into account the fact that the contributions descending from the off-diagonal corrections cannot be underestimated in the correct computation of the dark radiation abundance. Thus, the inclusion of quantum corrections in the K{\"a}hler metric made it evident that logarithmic quantum effects enhance the contributions to dark radiation, where the effective number of neutrino species $\Delta N_{eff}$ is calculated for both case studies remains well below the upper astrophysical bounds. \par

Finally, a proposal for non-thermal dark matter was presented and the two main production mechanisms were studied. The branching scenario is favored by considering exponentially suppressed fluxes, where the reheating temperature is well below the $\text{Tev}$ scale. In this scenario, the dark matter mass lays at the $ \mathcal{O}(\text{GeV})$ scale. Additionally, branching scenario and the late-time decays of moduli can accommodate a baryogenesis mechanism where the Baryon-Dark Matter Coincidence can be safely explained. The branching ratio of the N species particles and the CP asymmetry due to the decays to the particles of the observable sector are correlated to the dynamics of the moduli, through the reheating temperature. This way dark matter abundance and the baryon asymmetry can be constrained and derived from the requirements that a dS vacuum exists. In terms of the annihilation scenario, reheating temperatures can be regarded as high-scale scenario, where the mass of dark matter candidates is superheavy at scales of $\mathcal{O}(10^{12}) \text{Gev}$.
\par

Some future interesting investigations would be to extend this analysis to more complex Calabi-Yau spaces and observe the dynamics in K{\"a}hler moduli sector. Additionally, a possible embedding of an inflationary model in accordance with the above analysis would be beneficial in order to clarify some ambiguities in the early universe, like the connection of dark sector dynamics to the inflationary observable quantities. From a phenomenological point of view, it would be interesting to understand the implications and the experimental signatures of the possible superheavy dark matter candidates, since various existing and future experiments, such as Ice Cube \cite{ 2008.04323} and RNO-G \cite{2010.12279}, are searching for state-of-the-art methods to probe the nature of dark matter. We retain all these questions for future work.

\section*{Acknowledgements}
The author would like to thank Emeritus Professor George K. Leontaris and Athanasios Karozas for numerous discussions regarding the nature of D-terms and for providing a crucial read and comments on the final draft of this paper.

\appendix
\section{Appendix}

\subsection{Mixing matrix}

Given the case studies discussed in section 3., we are going to characterize the mixing between the moduli and the normalized fields $\phi_i$, based on the mass matrix in equation \eqref{MassMatrix}. Following the procedure given in \cite{2203.08833,Cicoli:2010ha}, the mixing matrix $P_{ij}$ for the two cases can be written as:

\begin{align}
\begin{pmatrix}
\tau_1 \\
\tau_2 \\
\tau_3
\end{pmatrix}= \begin{pmatrix}
\\
\vec{u}_1\\
~
\end{pmatrix}\phi_1 + \begin{pmatrix}
\\
\vec{u}_2\\
~
\end{pmatrix}\phi_2 + \begin{pmatrix}
\\
\vec{u}_3\\
~
\end{pmatrix}\phi_3,\;\; P_{ij}=\begin{pmatrix}
    & \vec{u}_1 &\\
     & \vec{u}_2 & \\
     & \vec{u}_3 &
\end{pmatrix}^T~.
\end{align}
For the derivation of the corresponding eigenvectors can be derived by following the recipe:

\begin{align}
    M^2_{ij} \vec{u}_i=m^2_i \vec{u}_i, \;\;\; \vec{u}_i^T \cdot \mathcal{K} \cdot \vec{u}_j=\delta_{ij},
\end{align}

where the two components of each eigenvector will be defined by the first relation, while the normalization condition will fix the latter component. Consequently, for each case discussed in section 3. with respect to the scale of integer fluxes, the mixing of the moduli can be approximated to:

\begin{equation}
    P_{ij}^{A)}\sim \begin{pmatrix}
\ \frac{3\mathcal{W}_0 \alpha \beta}{\mathcal{V}} & \frac{2d^{5/3}\delta}{\epsilon} & \frac{2d^{5/3}\delta}{\epsilon}\\
\frac{8\beta}{\mathcal{W}_0} & -(-\mathcal{W}_0)^{1/4} & \frac{32\beta}{5\mathcal{W}_0}\\
\frac{3\mathcal{W}_0\alpha \beta}{5d \mathcal{V}} & \frac{-8\beta}{\mathcal{W}_0} & \frac{9\mathcal{W}_0 \alpha \beta}{5d \mathcal{V}}
\end{pmatrix},\quad  P_{ij}^{B)}\sim 
\begin{pmatrix}
\frac{3\mathcal{W}_0 \alpha \beta}{10} & -\frac{10 d^{5/3}\delta}{\gamma} & -\frac{3\mathcal{W}_0 \alpha \beta}{\mathcal{V}}  \\
-\frac{32\beta}{5\mathcal{W}_0} & 1/2 & \frac{32\beta}{5\mathcal{W}_0}\\
-\frac{19\mathcal{W}_0\alpha \beta}{8d \mathcal{V}} & \frac{2d^{2/3}\delta}{\gamma} & -\frac{19\mathcal{W}_0 \alpha \beta}{8d \mathcal{V}}
\end{pmatrix},\label{mixing}
\end{equation}

where the variables $\alpha, \beta, \gamma, \delta, \epsilon$ are defined as:

\begin{align}
    \alpha=(-28\eta +3\xi+6\eta \log(\mathcal{V}))&,\quad \beta=(\frac{d\mathcal{V}}{-55\eta+6\xi})^{1/2},\quad \gamma=-3-2q+8(d^2\mathcal{V})^{1/3},\notag \\
    &\delta=(1+2q)\mathcal{V}^{1/3},\quad \epsilon=(3+2q)\mathcal{W}_0^2~.
\end{align}

The corresponding eigenvalues for both cases are given below:

\begin{align}
   m^{2\; }_{\phi_i}\cong \big(Tr[M^2],Tr[M^2], \dfrac{Det[M^2]}{Tr[M^2]^2}\big)~.
\end{align}

Now, the crucial next step is to define which normalized field mainly describes each one of the geometric moduli. Starting from the first case $A)$, one can easily observe the dependence of the textures on the integer fluxes $\mathcal{W}_0$. This observation is important because, in this case, this parameter is exponentially small, providing a guide on how to determine the leading contributions in the mixing matrix. Thus, we compare the ratios between the aforementioned entries of the matrix.

\begin{align}
    \frac{P_{12}^{A)}}{P_{11}^{A)}}= \frac{2 d^{1/3} (\mathcal{W}_0^2 \eta  q)^{1/3}}{6 \eta \mathcal{W}_0^2} \frac{1}{\beta \epsilon},\quad 
    \frac{P_{22}^{A)}}{P_{21}^{A)}}=\frac{(-\mathcal{W}_0)^{5/4}}{8\beta},\quad \frac{P_{23}^{A)}}{P_{31}^{A)}}=\frac{40 q}{6(2q+1)}
\end{align}

Given the scale of the parameters $\eta \sim \mathcal{O}(0.1),\;\xi \sim \mathcal{O}(1), \;\mathcal{W}_0 \ll \mathcal{O}(1), \; d\ll \mathcal{O}(10^{-3})$, we can deduce the following mixing between the moduli and the normalized fields at the leading order:

\begin{align}
    &\tau_1 \cong  P_{12}^{A)} (\phi_2 +\phi_3) , \notag \\
    &\tau_2 \cong   P_{21}^{A)} \phi_1 +  P_{23}^{A)} \phi_3,\notag \\
    &\tau_3 \cong P_{32}^{A)} \phi_2 ~.
\end{align}

From the above, it is remarkable that in the regime of exponentially small fluxes $|\mathcal{W}_0|\ll 1$, there exists a geometric separation between the world volumes. The overall volume is given mainly by $\phi_2$, while the transverse directions are approximately independent of this field. Even in the case of $\mathcal{V}\gg 1$, the uplift parameter d will compensate the suppression of the mixing, providing this nice geometric result.  

The second case $B)$ of mixing following the same reasoning results in qualitatively the same mixing. It is important to examine the ratios of the entries:

\begin{align}
    \frac{P_{12}^{B)}}{P_{11}^{B)}}=& \frac{25 q ({\frac{d \eta  q}{\mathcal{W}_0^2}})^{1/3}}{6 \mathcal{W}_0^2} \frac{1}{\beta \gamma},\quad 
    \frac{P_{22}^{B)}}{P_{21}^{B)}}=\frac{5\mathcal{W}_0}{64\beta},\quad \frac{P_{23}^{B)}}{P_{31}^{B)}}=\frac{16 q ({\frac{d \eta  q}{\mathcal{W}_0^2}})^{1/3}}{38 \mathcal{W}_0^2} \frac{1}{\beta \gamma},\notag \\
    & \beta= \sqrt{\frac{q \mathcal{W}_0^2}{-55+6\xi/\eta}},\quad \gamma= (3+2q -8 (\eta q d \mathcal{W}_0)^{1/3})~.
\end{align}

Given the scale of the parameters $\eta \sim \mathcal{O}(0.1),\;\xi \sim \mathcal{O}(1), \;\mathcal{W}_0 \sim \mathcal{O}(1), \; d\sim \mathcal{O}(10^{-4},10^{-3})$, we can deduce the following mixing between the moduli and the normalized fields at the leading order:

\begin{align}
    &\tau_1 \cong  P_{12}^{B)} \phi_2 , \notag \\
    &\tau_2 \cong   P_{21}^{B)} \phi_1 +  P_{23}^{B)} \phi_3,\notag \\
    &\tau_3 \cong P_{32}^{B)} \phi_2 ~.
\end{align}

Again in this case, the overall volume $\mathcal{V}$ is given by a single normalized field $\phi_2$. It is remarkable that geometric separation is a generic feature shared by this compactified space, where this fact is unraveled only after the process of finding the correct eigenvectors of the system. This feature was not given much attention in previous works \cite{1810.05060}, where the inflation scenario was studied as a multi-field system. This could be avoided after picking the appropriate scale for the fluxes $\mathcal{W}_0$. A final remark is that the computations given in this appendix will also be used in section 4., where the coupling of the normalized fields to the axions and Higgses will be calculated. 

\subsection{A case with suppressed loop corrections}
In the main body of the text, we argue that loop corrections are necessary for the correct formulation of the predictions of the model for dark radiation and dark matter. In this short section, we show that decreasing the effects of the string loop corrections inverts the hierarchy of the moduli masses, leading to a different longest-lived particle. This, in turn, results in the overproduction of dark radiation and eliminates the solution for the baryon–dark matter coincidence. Therefore, we investigate the case $\mathcal{W}_0 \ll 1$, since this condition is required for the baryon–dark matter coincidence and for achieving low reheating temperatures.\par
For the following set of parameters, the $q$ parameter which is required for having a dS vacuum is given by:
\begin{equation}
    |\mathcal{W}_0| = 10^{-4},\; \eta=-0.15,\; \xi=2, \;q=-1.03, \; \mathcal{V}_{min}= 1.2\times 10^5, \; d=1.3\times 10^{-14}~.
\end{equation}

Taking into account the mixing factor of equation \eqref{mixing}, the eigenvalues in this limit are given by:
\begin{align}
    \dfrac{Tr[M^2]}{\frac{Det[M^2]}{Tr[M^2]^2}}<1\Rightarrow m_{\tau_3}^2> m_{\tau_1}^2\cong m_{\tau_2}^2~.
\end{align}
This case opposes the scenario discussed in the current model. Due to the different hierarchy in the masses, the longest-lived particle will also change. From the matrix $P_{ij}^{A)}$, we can compute the new decay rate of the longest-lived particle:
\begin{align}
    \Gamma_{\phi_2 \rightarrow H \bar{H}} =\frac{432 \sqrt{2} d^{29/6} (2 q+1)^2 Z^2}{\pi  (2 q+3)^2 \mathcal{V}^{7/3} \mathcal{W}_0^4} \ll \Gamma_{\phi_1}, \Gamma_{\phi_3}~.
\end{align}
where we used the mixing factor $P_{12}^{A)}$ and the fact that the normalized field descends from $\tau_1$, since it is the lightest. Now, in order to compute the decay rates of moduli to axions, we use now the matrix $\partial_ {\tau_1}\mathcal{K}_{ij}$ instead of $\partial_ {\tau_3}\mathcal{K}_{ij}$.  The relevant Lagrangian part for the decays to axions is written as:
\begin{align}
     \mathcal{L} \supset  \mathcal{K}_{1np} P^{\tau}_{12} P^{c}_{nj} P^{c}_{pk} m^2_{\phi_2} \phi_2 \alpha_j \alpha_k~,
\end{align}
Assuming the resulting axions are $\alpha_3$, the decay rate can be approximated to:
\begin{align}
    \Gamma_{\phi_2 \rightarrow \alpha_3 \alpha_3}\cong \frac{54 \sqrt{2} d^{5/6} \lambda _{33}^4 (2 q+1)^2}{\pi  (2 q+3)^2 \mathcal{V}^{19/3} \mathcal{W}_0^4} + \eta d f(\lambda_{ij}),
\end{align}
where the first term corresponds to the tree level contribution and the second parametrizes the contributions from the off-diagonal corrections. The first observation from the above result is that in this case, the tree level contributions are not suppressed by $d$ compared to equation \eqref{axion1}. Substituting the parameters $\alpha, \;\beta$, we can compute the contribution to dark radiation as:
\begin{equation}
    \Delta N_{eff} \cong \frac{\lambda _{33}^4}{8 d^4 \mathcal{V}^4 Z^2} \cong \dfrac{  \lambda _{33}^4}{8\eta^4 \mathcal{W}_0^8q^4 Z^2} \gg 1~,
\end{equation}
where in the last step we used $q=(d\mathcal{V}_{min})/(\eta \mathcal{W}_0^2)$. The reheating temperature is given by:
\begin{equation}
    T_{rh} \cong \frac{d^{29/12}  Z \left| 2 q+1\right| }{\mathcal{V}^{7/6} \mathcal{W}_0^2 \left| 2 q+3\right| } M_p << \mathcal{O}(MeV)~.
\end{equation}
Given the two calculations above, we can deduce that the model with suppressed $\eta$ is inconsistent with the cosmological observations.

\subsection{Decay rate formulas}

For the derivation of the decay rates given in the main body of this paper, we used the standard formula:

\begin{align}
\Gamma=\frac{1}{S} \int \frac{|M|^2}{2E} dLIPS,
\end{align}

where the $d_{LIPS}$ is the element of the Lorentz invariant phase space and $S$ is the symmetry factor. The decaying particle's energy is parametrized by $E$. There are two possible decay channels, which can be written as:

\begin{align}
\mathcal{L} \supset g \phi_i \psi^2+ g \phi_i \psi \chi~.
\end{align}

In the above equation, we have assumed that the mass of $\phi_i$  is much heavier than $\psi,\chi$. The symmetry factor and the matrix element for the first case is given $S=2$ and $|M|^2=4g^2$, while for the latter one is summarized to $S=1, |M|^2=g^2$. The corresponding coupling $g$ in each process will be read by the Lagrangian terms, so the two decay rates are evaluated to be:

\begin{align}
\Gamma_{\phi_i \rightarrow \psi \psi}=\frac{g^2}{8\pi m_{\phi_i}},\;\; \Gamma_{\phi_i \rightarrow \psi \chi}=\frac{g^2}{16\pi m_{\phi_i}}~.
\end{align}

\subsection{D-terms}

In this appendix, we provide a detailed stabilization using the generic D-terms formula, following the work of \cite{2203.03362,Bera:2024zsk, Ahmed:2022dhc}, and we focus on finding the relevance of the derived vacuum with our approximation in equation \eqref{Dterm}. Starting from the generic formula of the D-terms \eqref{GenD}, it can be expanded to:

\begin{align}
     V_D \cong \sum_{i=3}^3 
    \big[\dfrac{1}{\tau_i}\big(\sum_{i\neq j}Q_{ij} \partial_{T_j} \mathcal{K}\big)^2] = \frac{d_1}{\tau _1}\left(\frac{Q_{12}}{\tau _2}+\frac{Q_{13}}{\tau _3}\right){}^2 + \frac{d_2 }{\tau _2} \left(\frac{Q_{21}}{\tau _1}+\frac{Q_{23}}{\tau _3}\right){}^2 + \frac{d_3 }{\tau _3} \left(\frac{Q_{31}}{\tau _1}+\frac{Q_{32}}{\tau _2}\right){}^2 ~.
\end{align}

The global embedding of this toy model has been analyzed in \cite{Bera:2024zsk}. In this work, we need to stabilize two moduli by the D-terms, which fact is of particular importance for embedding consistent inflationary paradigms in such string scenarios \cite{2203.03362,Bera:2024zsk, Ahmed:2022dhc}. In order to do so, we are going to assume that the charges $Q_{ij}$ obey to the following relations:

\begin{equation}
    Q_{12}=Q_{23}=Q_{31},  \quad Q_{21}=Q_{13}=Q_{32}, \quad Q_{12}\neq Q_{21}~.
\end{equation}

In addition to that, we could in general assume that $Q_{12}=1, Q_{21}=0$, since this could significantly simplify the form of the moduli eigenvalues and eigenvectors, helping us to study the qualitative behavior of dark radiation in a stabilized dS vacuum. Also, since we would like to compare this new vacuum with the vacuum presented in the main body of the paper, we are going to redefine the $\tau_1,\tau_2$ moduli as $\tau_1^{\prime},\tau_2^{\prime}$.

\begin{equation}
    V_D = \frac{d_1}{\tau _2^{\prime 2} \tau _1^{\prime}}+\frac{d_2 \tau _2^{\prime} \tau _1^{\prime 2}}{\mathcal{V}^4}+\frac{d_3 \tau _2^{\prime}}{\tau _1^{\prime} \mathcal{V}^2}~.
\end{equation}

Minimizing with respect to $\tau_1^{\prime}$, $\tau_2^{\prime}$ and $\mathcal{V}$, we have the following minima:

\begin{align}
    \tau_1^{\prime} = &(\dfrac{d_3}{d_2})^{1/3}\mathcal{V}^{2/3}, \quad \tau_2^{\prime} = (\dfrac{d_1}{d_3})^{1/3}\mathcal{V}^{2/3}, \notag \\
    & \mathcal{V}_{min} = \frac{3 \eta  \mathcal{W}_0^2 W_{0/-1}\left(\frac{2 d e^{\frac{13}{3}-\frac{\xi }{2 \eta }}}{3 \eta  \mathcal{W}_0^2}\right)}{2 d}~.\label{newvac}
\end{align}

After applying the minimal values of $\tau_1^{\prime}, \tau_2^{\prime}$ at the D-terms, we get the expected formula:

\begin{align}
    V_{D}= \dfrac{3d}{\mathcal{V}^2},\quad  d=(d_1 d_2 d_3)^{1/3}~.
\end{align}

Readily, one can see that we have arrived to the exact same minimal value for the volume modulus $\mathcal{V}_{min}$. Taking into account the derived minima in equation \eqref{mins} and the minima derived from the generic form \eqref{newvac}, it readily found that they are related up to a scaling in the uplifting parameter $d$.

\begin{align}
    &\tau_1 = \tau_1^{\prime} \dfrac{(d_1 d_2)^{2/9}}{d_3^{4/9}}, \quad \tau_2 = \tau_2^{\prime} \dfrac{(d_2 d_3)^{2/9}}{d_1^{4/9}},  \notag\\
    &\tau_1= d^{2/3} \tau_1^{\prime}, \quad \tau_2 = \dfrac{\tau_2^{\prime}}{d^{4/3}}~.
\end{align}

In the last step we have used $d_1= d^{3}/(d_2d_3), \;d_2=1,\;d_3=1$. Consequently, we can deduce that this equivalence of the vacua does not spoil the analysis with respect to the observable quantities in this model, since the scaling (and the minimal value) with respect to the compactified volume $\mathcal{V}$ is the same.

\end{document}